\documentclass[%
preprint,
tighten,
amsmath,
amssymb,
aps,
twocolumn,
]{aastex631}

\usepackage{graphicx}
\usepackage{dcolumn}
\usepackage{bm}
\usepackage{color}
\usepackage{natbib}
\usepackage{comment}
\usepackage{verbatim}

\newcommand{\oversnp}{6}

\newcommand{\ds}{4}
\newcommand{\msig}{$M_{\mathrm{BH}}-\sigma_\star$}
\newcommand{\mlum}{$M_{\mathrm{BH}}-L_{\mathrm{bul}}$}
\newcommand{\mmass}{$M_{\mathrm{BH}}-M_{\mathrm{bul}}$}

\newcommand{\mbh}{$M_{\mathrm{BH}}$}
\newcommand{\inc}{\ensuremath{i}}
\newcommand{\ml}{$M/L_\mathrm{H}$}
\newcommand{\pa}{\ensuremath{\Gamma}}
\newcommand{\vsys}{$v_{\text{sys}}$}
\newcommand{\xloc}{$x_0$}
\newcommand{\yloc}{$y_0$}
\newcommand{\fw}{$f_0$}
\newcommand{\sig}{$\sigma_0$}

\newcommand{\kms}{km s$^{-1}$}

\newcommand{\mbhmed}{7.26}
\newcommand{\mbhups}{0.43}
\newcommand{\mbhdowns}{0.48}
\newcommand{\mbhupsss}{1.36}
\newcommand{\mbhdownsss}{1.74}

\newcommand{\mbhupsys}{0.55} 
\newcommand{\mbhdownsys}{1.00} 
\newcommand{\mlmed}{1.34}
\newcommand{\mlups}{0.01}
\newcommand{\mldowns}{0.01}
\newcommand{\mlupsss}{0.03}
\newcommand{\mldownsss}{0.03}

\newcommand{\mldownsys}{0.02}

\newcommand{\mbhfullerrngc}{\mbh\ $= (\mbhmed ^{+\mbhups}_{-\mbhdowns}$ [$1\sigma$ stat] $^{+\mbhupsys}_{-\mbhdownsys}$ [sys])$\times 10^8$ $M_\odot$}
\newcommand{\mbhfullerrforabstractngc}{\mbh\ $= (\mbhmed^{+\mbhups}_{-\mbhdowns}$ [$1\sigma$ statistical] $^{+\mbhupsys}_{-\mbhdownsys}$ [systematic])$\times 10^8$ $M_\odot$}
\newcommand{\mbhbothsigngc}{\mbh\ $= (\mbhmed ^{+\mbhups}_{-\mbhdowns}$ [stat, $1\sigma$] $^{+\mbhupsss}_{-\mbhdownsss}$ [stat, $3\sigma$] $^{+\mbhupsys}_{-\mbhdownsys}$ [sys])$\times 10^8$ $M_\odot$}
\newcommand{\mlabstract}{$M/L_H = \mlmed\pm\mldowns$ [$1\sigma$ statistical] $\pm\mldownsys$ [systematic] $M_\odot/L_\odot$}  
\newcommand{\mlbothsig}{$M/L_H = \mlmed\pm\mldowns$ [stat, $1\sigma$] $\pm\mldownsss$ [stat, $3\sigma$] $\pm\mldownsys$ [sys] $M_\odot/L_\odot$} 
\newcommand{\mlfullerr}{$M/L_H = \mlmed\pm\mldowns$ [$1\sigma$ stat] $\pm\mldownsys$ [sys] $M_\odot/L_\odot$} 

\begin{document}

\shortauthors{Cohn et al.}

\title{Modeling ALMA Observations of the Warped Molecular Gas Disk in the Red Nugget Relic Galaxy NGC 384}

\author[0000-0003-1420-6037]{Jonathan H. Cohn}
\affiliation{George P. and Cynthia W. Mitchell Institute for Fundamental Physics and Astronomy, Department of Physics \& Astronomy, Texas A\&M University, 4242 TAMU, College Station, TX 77843, USA}
\affiliation{Department of Physics and Astronomy, Dartmouth College, 6127 Wilder Laboratory, Hanover, NH 03755, USA}

\author[0000-0003-4612-5186]{Maeve Curliss}
\affiliation{George P. and Cynthia W. Mitchell Institute for Fundamental Physics and Astronomy, Department of Physics \& Astronomy, Texas A\&M University, 4242 TAMU, College Station, TX 77843, USA}

\author[0000-0002-1881-5908]{Jonelle L. Walsh}
\affiliation{George P. and Cynthia W. Mitchell Institute for Fundamental Physics and Astronomy, Department of Physics \& Astronomy, Texas A\&M University, 4242 TAMU, College Station, TX 77843, USA}

\author[0000-0003-2632-8875]{Kyle M. Kabasares}
\affiliation{Ames Research Center, National Aeronautics and Space Administration, Moffett Field, CA 94035, USA}
\affiliation{Bay Area Environmental Research Institute, Ames Research Center, Moffett Field, CA 94035, USA}
\affiliation{Department of Physics and Astronomy, 4129 Frederick Reines Hall, University of California, Irvine, CA, 92697-4575, USA}

\author[0000-0001-6301-570X]{Benjamin D. Boizelle}
\affiliation{Department of Physics and Astronomy, N284 ESC, Brigham Young University, Provo, UT 84602, USA}
\affiliation{George P. and Cynthia W. Mitchell Institute for Fundamental Physics and Astronomy, Department of Physics \& Astronomy, Texas A\&M University, 4242 TAMU, College Station, TX 77843, USA}

\author[0000-0002-3026-0562]{Aaron J. Barth}
\affiliation{Department of Physics and Astronomy, 4129 Frederick Reines Hall, University of California, Irvine, CA 92697-4575, USA}

\author[0000-0002-8433-8185]{Karl Gebhardt}
\affiliation{Department of Astronomy, The University of Texas at Austin, 2515 Speedway, Stop C1400, Austin, TX 78712, USA}

\author[0000-0002-1146-0198]{Kayhan G\"{u}ltekin}
\affiliation{Department of Astronomy, University of Michigan, 1085 S. University Ave., Ann Arbor, MI 48109, USA}

\author[0000-0002-3202-9487]{David A. Buote}
\affiliation{Department of Physics and Astronomy, 4129 Frederick Reines Hall, University of California, Irvine, CA 92697-4575, USA}

\author[0000-0003-2511-2060]{Jeremy Darling}
\affiliation{Center for Astrophysics and Space Astronomy, Department of Astrophysical and Planetary Sciences, University of Colorado, 389 UCB, Boulder, CO 80309-0389, USA}

\author[0000-0002-7892-396X]{Andrew J. Baker}
\affiliation{Department of Physics and Astronomy, Rutgers, the State University of New Jersey, 136 Frelinghuysen Road Piscataway, NJ 08854-8019, USA}
\affiliation{Department of Physics and Astronomy, University of the Western Cape, Robert Sobukwe Road, Bellville 7535, South Africa}

\author[0000-0001-6947-5846]{Luis C. Ho}
\affiliation{Kavli Institute for Astronomy and Astrophysics, Peking University, Beijing 100871, China; Department of Astronomy, School of Physics, Peking University, Beijing 100871, China}

\correspondingauthor{Jonathan H. Cohn}
\email{jonathan.cohn@dartmouth.edu}

\begin{abstract}

We present 0$\farcs{22}$-resolution CO(2$-$1) observations of the circumnuclear gas disk in the local compact galaxy NGC 384 with the Atacama Large Millimeter/submillimeter Array (ALMA).
While the majority of the disk displays regular rotation with projected velocities rising to $370$ \kms, the inner $\sim$0\farcs{5} exhibits a kinematic twist.
We develop warped disk gas-dynamical models to account for this twist, fit those models to the ALMA data cube, and find a stellar mass-to-light ratio in the $H$-band of \mlabstract\ and a supermassive black hole (BH) mass (\mbh) of \mbhfullerrforabstractngc.
In contrast to most previous dynamical \mbh\ measurements in local compact galaxies, which typically found over-massive BHs compared to the local BH mass$-$bulge luminosity and BH mass$-$bulge mass relations, NGC 384 lies within the scatter of those scaling relations.
NGC 384 and other local compact galaxies are likely relics of $z\sim2$ red nuggets, and over-massive BHs in these relics indicate BH growth may conclude before the host galaxy stars have finished assembly.
Our NGC 384 results may challenge this evolutionary picture, suggesting there may be increased scatter in the scaling relations than previously thought.
However, this scatter could be inflated by systematic differences between stellar- and gas-dynamical measurement methods, motivating direct comparisons between the methods for NGC 384 and the other compact galaxies in the sample.
\end{abstract}

\section{\label{intro}Introduction}

Dynamical supermassive black hole (BH) detections have been made in over 100 nearby galaxies, with stellar dynamics accounting for the vast majority of these measurements \citep{Saglia2016}.
In recent years, the advent of the Atacama Large Millimeter/sub-millimeter Array (ALMA), with its improved angular resolution and sensitivity compared to previous mm/sub-mm observatories, has led to a significant increase in molecular gas-dynamical BH mass (\mbh) measurements in the literature (e.g., \citealt{Barth2016a,Davis2017,Boizelle2019,Boizelle2021,Nagai2019,North2019,Smith2021,Cohn2021,Cohn2023,Kabasares2022,Nguyen2022,Ruffa2023,KabasaresCohn2024,Dominiak2024a,Dominiak2024b}).
The cold molecular gas detected with ALMA is a more reliable tracer of the gravitational potential around BHs than warm H$_2$ molecular gas or ionized gas, which often display more turbulent motion (e.g., \citealt{MarelBosch1998,Barth2001, Wilman2005,Neumayer2007,Seth2010,Walsh2010,Walsh2013,Scharwachter2013}).

Via these dynamical measurements, BH masses have been found to correlate with large-scale properties of their host galaxies, including stellar velocity dispersion ($\sigma_\star$), bulge mass ($M_\mathrm{bul}$), and bulge luminosity ($L_\mathrm{bul}$; e.g., \citealt{Kormendy1995,Ferrarese2000,Gebhardt2000,Marconi2003,Gultekin2009,Kormendy2013}), indicating BHs and their hosts co-evolve.
However, this BH$-$host galaxy co-evolution is not well understood, in part because the galaxies in which dynamical \mbh\ measurements have been made are not fully representative of the population of galaxies in the Universe.
Furthermore, dynamical \mbh\ measurements are not generally possible beyond the local Universe, complicating attempts to study the cosmic evolution of the BH scaling relations.

Here, we study the circumnuclear molecular gas disk in NGC 384 with ALMA.
NGC 384 is a member of a sample of local, massive, and compact early-type galaxies (ETGs) found through the Hobby-Eberly Telescope Massive Galaxy Survey (HETMGS; \citealt{Bosch2015}).
\citet{Yildirim2017} studied the sample's stellar kinematics and photometric properties.
These galaxies have large stellar velocity dispersions \citep{Yildirim2017}, indicating they likely have large BHs (\mbh\ $\lesssim6\times10^9\ M_\odot$; \citealt{Kormendy2013,Saglia2016}), populating the poorly-sampled high-mass end of the scaling relations.
They are also massive (stellar masses $M_\star \sim 5.5\times10^{10} - 3.8\times10^{11}\ M_\odot$) and compact (effective radii $r_\mathrm{e} \sim 0.7-3.1$ kpc), falling on the redshift ($z$) $\sim2$ size$-$mass relation, despite lying at distances within $\sim$100 Mpc, i.e., at $z\sim0$ \citep{Yildirim2017}.
These systems are fast rotators with cuspy surface brightness profiles, flattened, disk-like shapes, and no evidence in their stellar orbital distributions for major mergers since $z\sim2$ \citep{Yildirim2017}.
They exhibit uniform, old stellar ages ($\gtrsim$10 Gyr) over several effective radii \citep{Martin2015,Mateu2017,Yildirim2017} and elevated fractions of red globular clusters \citep{Beasley2018,Kang2021}.
Individual objects in the sample have also been shown to have highly concentrated dark matter halos \citep{Buote2018,Buote2019}.

These local compact galaxies are very different from the brightest cluster galaxies and giant ETGs in the local Universe that typically host the highest-mass BHs.
The latter, more typical massive ETGs are commonly thought to evolve from quiescent red nugget galaxies observed at $z\sim2$, growing through accretion and minor/intermediate dry mergers without significant BH growth (e.g., \citealt{Dokkum2010,Hilz2013}).
In contrast, local compact galaxies like NGC 384 are likely passively evolved relics of the red nugget galaxies \citep{Trujillo2014,Yildirim2017}.
Thus, if the passively evolved relics of red nugget galaxies tend to host over-massive BHs, the BHs of common massive local ellipticals may have finished growing by $z\sim2$ (e.g., \citealt{Mateu2015}).

Aside from these relic galaxies, most of our knowledge of BH$-$host galaxy co-evolution at high $z$ stems from single epoch Active Galactic Nucleus (AGN) measurements (e.g., \citealt{Izumi2019,Pensabene2020,Larson2023,Maiolino2023,Bogdan2024}).
In particular, recent single epoch \mbh\ determinations from AGN have pointed to BHs that are over-massive at high $z$ compared to the local scaling relations \citep{Pacucci2023}.
However, there is a systematic factor of two uncertainty in the local reverberation mapping \mbh\ measurements to which single epoch measurements are anchored \citep{Shen2023}, and single epoch BH masses can be further overestimated by $\sim$0.3 dex when the diversity of quasar properties is not accounted for \citep{FonsecaAlvarez2020}.
As such, complementary methods to study high-$z$ BH growth are required.
The local compact relic galaxy sample thus presents a remarkable alternative view into the history of the BH scaling relations.

Black hole mass measurements already exist for five galaxies in this sample, including three using stellar dynamics and two using molecular gas dynamics from ALMA.
The measurements from stellar dynamics (in NGC 1277, NGC 1271, and Mrk 1216; \citealt{Bosch2012,Emsellem2013,Yildirim2015,Walsh2015,Walsh2016,Walsh2017,Graham2016a,Krajnovic2018}) find BH masses consistent with the \msig\ relation but over-massive compared to the \mlum\ and \mmass\ relations by an order of magnitude, even when conservatively using the total stellar luminosities and total stellar masses of the galaxies, rather than their uncertain bulge values (e.g., \citealt{Graham2016b,SavorgnanGraham2016}).
\citet{Scharwachter2016} observed molecular gas in NGC 1277 with the IRAM Plateau de Bure Interferometer, finding an \mbh\ consistent with stellar-dynamical studies, albeit with significant uncertainties due to limited angular resolution.
At much higher angular resolution, one ALMA measurement (PGC 11179; \citealt{Cohn2023}) echoed the behavior of the over-massive stellar-dynamical results, but the other (UGC 2698; \citealt{Cohn2021}) was consistent with all three relations.
However, UGC 2698 may have undergone some stellar growth since $z\sim2$ \citep{Yildirim2017}, evolving it toward the local BH scaling relations.

Although these results could be explained by greater than expected intrinsic scatter in the scaling relations rather than systematically different growth histories, the apparent scatter may be inflated due to a possible systematic offset between stellar- and molecular gas-dynamical measurement methods \citep{Cohn2021}.
Thus, to decipher whether the local compact galaxies truly display evidence for BHs finishing their growth before their host galaxy finishes assembling stars, we require more BH mass measurements in the sample and direct comparisons between stellar- and gas-dynamical results for individual objects.

In this work, we measure the BH mass in NGC 384 with molecular gas dynamics, accounting for a kinematic twist in the gas disk.
We adopt an angular diameter distance of 55 Mpc to NGC 384, where 267 pc spans 1\arcsec, using a $\Lambda$CDM cosmology with $H_0 = 73$ km s$^{-1}$ Mpc$^{-1}$, $\Omega_M = 0.31$, and $\Omega_{\Lambda} = 0.69$.
We use the Virgo + Great Attractor + Shapley Supercluster infall model \citep{Mould2000} for the Hubble Flow distance in the NASA/IPAC Extragalactic Database\footnote[10]{\url{https://ned.ipac.caltech.edu/}}.
We note that \citet{North2019} made an \mbh\ measurement in NGC 383, which is in the same group as NGC 384, and assumed a distance of 66.6 Mpc to NGC 383 in the process.
The BH mass in our models scales linearly with the assumed distance to NGC 384.

The composition of the paper is presented below.
In \S\ref{observations}, we discuss the Hubble Space Telescope (HST) and ALMA observations of NGC 384.
In \S\ref{model}, we describe our dynamical model, warped disk methodology, and parameter optimization.
We present our model results in \S\ref{results}.
In \S\ref{discussion}, we discuss our resolution of the BH gravitational sphere of influence (SOI), compare the BH in NGC 384 to the scaling relations, and discuss the impact of our results on our understanding of the scaling relations and BH$-$host galaxy growth.
Finally, we present our conclusions in \S\ref{conclusions}.

\section{\label{observations}Observations}

Confident \mbh\ measurements with molecular gas arise from observations resolving circumnuclear gas extending within or near to the BH SOI.
Additionally, the host galaxy's stellar light profile must be characterized to measure the contribution of stars to the gravitational potential on small scales.
We therefore obtain HST Wide Field Camera 3 (WFC3) observations, as described in \S\ref{hst}.
We detail our ALMA observations in \S\ref{alma}.

\subsection{HST Observations\label{hst}}

Program GO-13050 (PI: van den Bosch) observed NGC 384 with HST WFC3 in the IR/F160W ($H$-band) and UVIS/F814W ($I$-band) filters on 2013 June 6.\footnote{Based on observations made with the NASA/ESA Hubble Space Telescope, obtained from the data archive at the Space Telescope Science Institute, which is operated by the Association of Universities for Research in Astronomy, Inc., under NASA contract NAS5-26555. The observations are associated with program \#13050.}
The $H$-band observations included three dithered full-array exposures, with four dithered short sub-array exposures.
These sub-array exposures avoid saturating the nucleus while better sampling the point spread function (PSF).
These images were processed in \citet{Yildirim2017} using the {\tt calwf3} pipeline and distortion-corrected, cleaned, and combined using {\tt AstroDrizzle} \citep{Gonzaga2012}.
Ultimately, the $H$-band image has an exposure time of 1354.5 s, a pixel scale of $0\farcs06$ pixel$^{-1}$, and a field of view (FOV) of $2\farcm7\times2\farcm6$.

\begin{figure*}
\includegraphics[width=\textwidth]{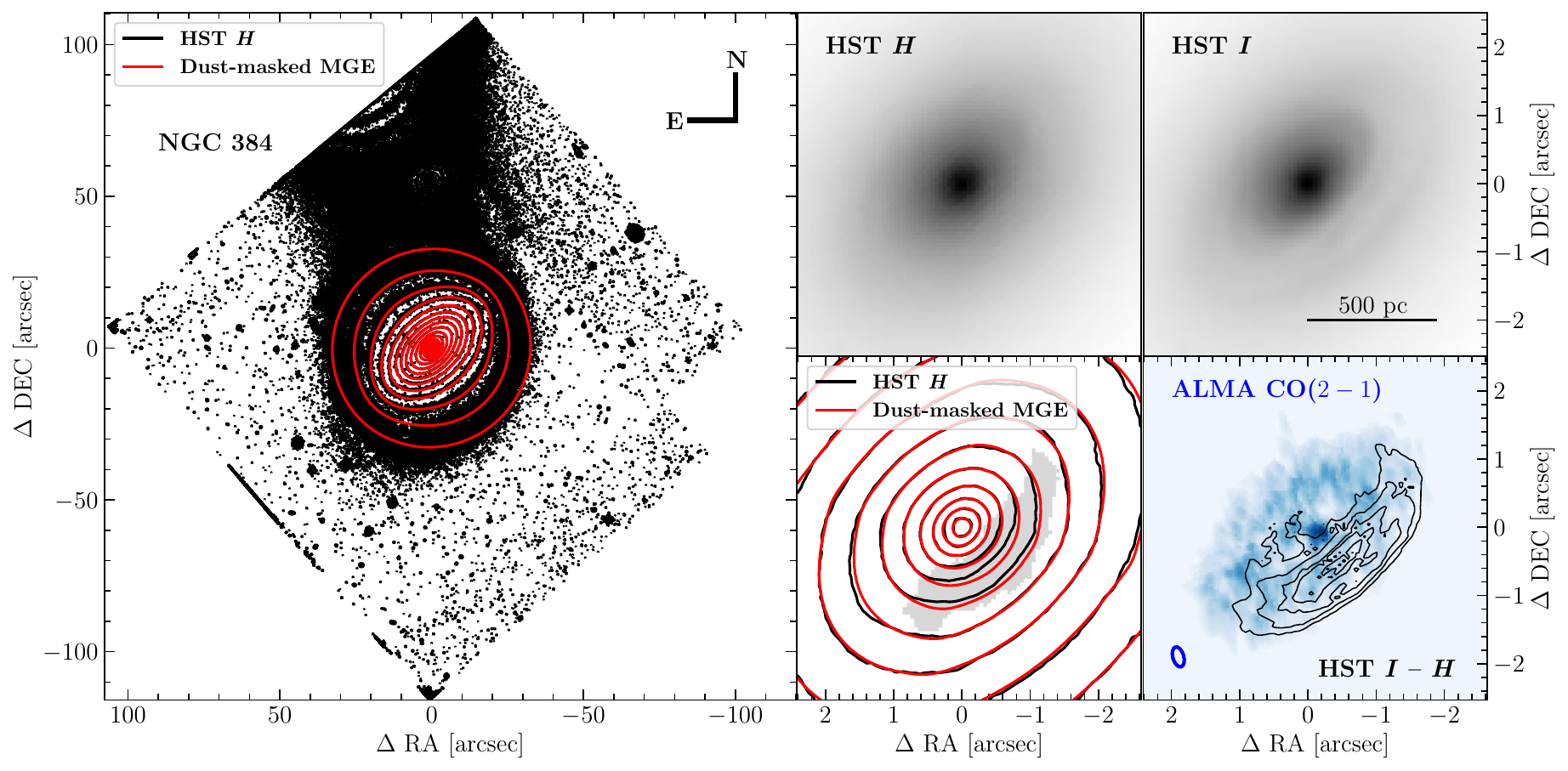}
\caption{(\textit{Left}) Contours of the HST F160W ($H$-band) image (black) with contours of the fiducial model's dust-masked MGE overlaid (red).
(\textit{Middle, top}) Central 5\arcsec\ of the HST $H$-band image.
(\textit{Right, top}) Central $5\arcsec$ of the HST F814W ($I$-band) image, which shows a dust disk to the southwest of the nucleus.
(\textit{Middle, bottom}) Central 5\arcsec\ of the HST $H$-band contours (black) and the dust-masked MGE (red), with gray shading showing the region that was masked during the MGE fit.
The asymmetry in the $H$-band contours in the dust-masked region points to dust attenuation, and the lack of asymmetry in the red contours indicates the dust-masked MGE accounts for this attenuation.
(\textit{Right, bottom}) ALMA CO($2-1$) emission (blue) with HST $I-H$ contours (black) overlaid, showing that the CO($2-1$) emission and dust disk are co-spatial and approximately the same size ($\sim$4$\farcs{2}$ in diameter).
The ellipse in the lower left of the panel corresponds to the ALMA synthesized beam.
}
\label{fig_hst_ngc}
\end{figure*}

Three dithered full-array exposures were also taken in the $I$-band.
The final $I$-band image, which has an exposure time of 482.0 s, is drizzled to the same scale as the $H$-band image and degraded to match the $H$-band image's resolution, so we can construct $I-H$ maps to characterize the dust disk in NGC 384.
These images are shown in Figure \ref{fig_hst_ngc}.

Dust is not clearly visible in the $H$-band image, but the $I$-band image shows a $\sim$4$\farcs{2}$-diameter circumnuclear dust disk or ring-like structure.
The median $I-H$ color is 1.7 mag, measured just beyond the disk region.
The maximum color excess $\Delta(I-H)$, which we find $\sim$0$\farcs{7}$ to the southwest of the nucleus, is $\sim$0.6 mag.
We use Vega-relative magnitudes throughout this work.

\subsubsection{Galaxy Surface Brightness Models\label{mge}}

In order to parameterize the stellar surface brightness profile of NGC 384, we fit two-dimensional (2D) Multi-Gaussian Expansions (MGEs) to the $H$-band image.
MGEs are capable of accurately reproducing the stellar surface brightness profiles of ETGs \citep{Emsellem1994,Cappellari2002}.
We complete an initial fit with a 2D regularized MGE \citep{Cappellari2002} and use that result as the initial guess for a 2D MGE fit in {\tt GALFIT} \citep{Peng2010}.
Gaussian position angles and centers are constrained to be identical for all components in {\tt GALFIT}.
The PSF, which comes from Tiny Tim \citep{Krist2004} and is drizzled and dithered identically to the image, is accounted for in the fit.

First, we fit an MGE to the $H$-band image, masking out foreground stars, other galaxies, and artifacts in the image.
This MGE is referred to as the ``original MGE'' throughout this paper.
The original MGE has ten components, with projected dispersions $\sigma^\prime$ ranging from $0\farcs{042}$ to $51\farcs{459}$, projected axis ratios $q^\prime$ between 0.581 and 1.000, and a PA of $136.732^\circ$ east of north.
The MGE is a good fit, with typical residuals of $\lesssim$5\%.

\begin{figure}
\includegraphics[width=0.47\textwidth]{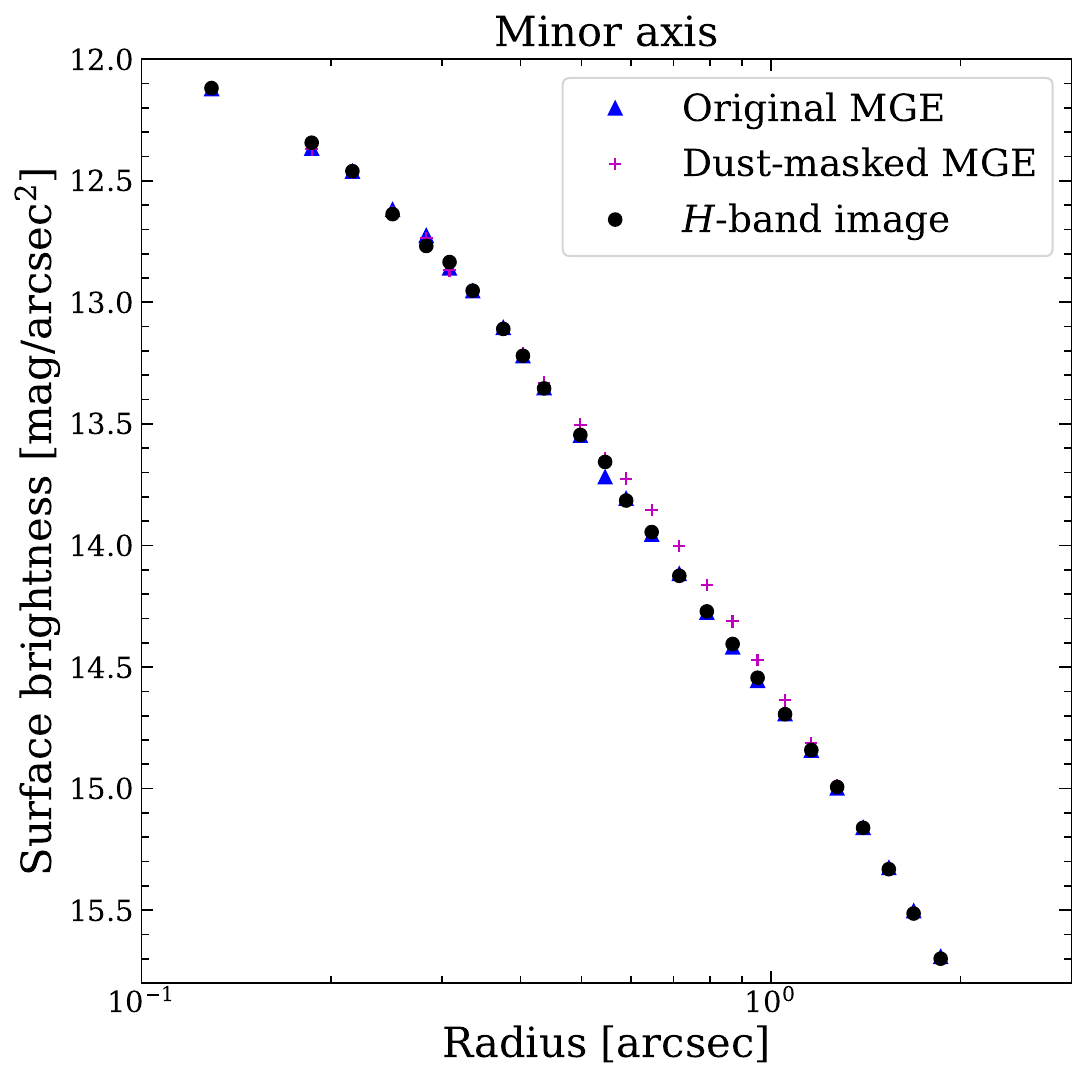}
\caption{Central surface brightness profiles as a function of projected radius along the minor axis.
Profiles are shown for the $H$-band image (black circles), original MGE (blue triangles), and fiducial dust-masked MGE (pink pluses).
As shown in Figure \ref{fig_hst_ngc}, the dust is located to the southwest of the nucleus along the minor axis.
The impact of the dust is visible as the difference between the surface brightness profiles in this figure at radii of $\sim$0\farcs{6}$-$1\arcsec.
}
\label{sbprof_ngcpgc}
\end{figure}

Next, we follow \citet{Cohn2021} and \citet{Cohn2023} in building an MGE with the dustiest portions of the $H$-band image masked out.
We start with a conservative color cut of $I-H = 2.05$ and fit an MGE to the resultant masked image in {\tt GALFIT}.
Then, we expand the mask to cover pixels with the greatest residuals near the nucleus and fit an MGE using this expanded mask.
We repeat this process iteratively until the residuals near the nucleus are $\lesssim$5\%.
The final mask and resultant MGE are called the ``dust mask'' and the ``dust-masked MGE'' throughout this paper.
We list the best-fit parameters of the dust-masked MGE in Table \ref{tab_mgengcpgc} and compare the dust-masked MGE and $H$-band image in Figure \ref{fig_hst_ngc}.
In Figure \ref{sbprof_ngcpgc}, we show the surface brightness profiles calculated along the minor axis within the inner $\sim$3\arcsec\ for the $H$-band image, the original MGE, and the dust-masked MGE.
The inner $\sim$0\farcs{2} of the galaxy appear mostly dust free.

Finally, we approximate a dust-corrected MGE by adjusting the $H$-band image before fitting in {\tt GALFIT}.
This method is described in more detail in \citet{Cohn2023}, but we summarize the process here.
First, we follow \citet{Viaene2017} and \citet{Boizelle2019} in assuming that the galaxy is oblate axysimmetric and that the dust lies in a thin disk in the galaxy inclined at the same angle as both the gas and the galaxy's stellar component.
We use an inclination of 57.3$^\circ$ from initial flat disk dynamical model results (see \S\ref{error_ngc}) to deproject the MGE, then calculate the fraction of light that originates behind versus in front of the dust disk at each pixel, taking the light originating behind the disk to be obscured with a simple screen extinction.
The model color excess at each pixel is calculated as a function of intrinsic dust extinction, $A_V$, using Equations 1 and 2 from \citet{Boizelle2019}.
To convert from $A_V$ to $A_H$ and $A_I$, we take the standard Galactic $R_V=3.1$ extinction curve \citep{Rieke1985}.
Following \citet{Cohn2023}, we do not attempt a full pixel-by-pixel correction.
The median $A_H$ inside the dust masked region is 0.4 mag, with a small standard deviation of 0.1 mag, and we correct the $H$-band image within the dust masked region by this median value.

We then use {\tt GALFIT} to fit a 2D Nuker model \citep{Faber1997} to the central $5\arcsec\times5\arcsec$ of the dust-masked $H$-band image, again accounting for the PSF.
The Nuker model consists of a double power-law with an inner slope $\gamma$, outer slope $\beta$, sharpness of transition $\alpha$, and break radius between the slopes $r_b$.
The fit converges to $\alpha=0.43$, $\beta=2.86$, $\gamma=0.00$, and $r_b=0.91\arcsec$ ($\sim$243 pc).
We re-fit this Nuker model to the 0.4 mag-adjusted $H$-band image in {\tt GALFIT}, holding $\beta$ and $r_b$ fixed, resulting in $\gamma=0.02$ and $\alpha = 0.44$.
Next, we replace the dust-masked region of the $H$-band image with the corresponding pixel values of the best-fit Nuker model, creating a dust-corrected image with a light distribution that varies smoothly.
We fit a final MGE to this dust-corrected image in {\tt GALFIT} and refer to this resultant MGE as the ``dust-corrected MGE'' throughout this paper.
The dust-corrected MGE consists of ten Gaussians, with $\sigma^\prime$ ranging between $0\farcs{047}$ and $45\farcs{015}$, $q^\prime$ between 0.581 and 1.000, a PA of 136.218$^\circ$ east of north, and residuals $\lesssim$3\%.
The major- and minor-axis surface brightness profiles of the dust-corrected MGE closely track those of the dust-masked MGE and are thus not included in Figure \ref{sbprof_ngcpgc}.
The MGEs we construct in this work follow a similar profile to the MGE presented for NGC 384 in \citet{Yildirim2017}.
Following \citet{Cohn2021} and \citet{Cohn2023}, we use the dust-masked MGE in the fiducial model and use the other two MGEs to assess the impact of our treatment of dust on the inferred \mbh.

\begin{deluxetable}{cccc}[ht]
\tabletypesize{\small}
\tablecaption{Dust-masked MGE parameters}
\tablewidth{0pt}
\tablehead{
\colhead{$j$} & 
\colhead{$\log_{10}(I_{H,j})$ [$L_{\odot}$ pc$^{-2}$]} & 
\colhead{$\sigma_j^\prime$ [arcsec]} & 
\colhead{$q_j^\prime$}
\\[-1.5ex]
\colhead{(1)} & 
\colhead{(2)} & 
\colhead{(3)} & 
\colhead{(4)}
}
\startdata
1 & 5.690 & 0.047 & 1.000 \\
2 & 5.072 & 0.222 & 0.717 \\
3 & 4.563 & 0.459 & 0.851 \\
4 & 4.196 & 1.032 & 0.705 \\
5 & 3.590 & 2.204 & 0.666 \\
6 & 3.254 & 5.165 & 0.581 \\
7 & 2.603 & 8.675 & 0.618 \\
8 & 1.893 & 14.679 & 0.974 \\
9 & 1.003 & 43.543 & 1.000 \\
10 & 0.977 & 44.858 & 1.000
\enddata
\begin{singlespace}
  \tablecomments{MGE parameters fit to the dust-masked HST $H$-band image of NGC 384 with {\tt GALFIT}.
  The MGE component is listed in Column (1).
  The central surface brightness of each component, given in Column (2), is calculated with an absolute Solar $H$-band magnitude of 3.37 mag \citep{Willmer2018} and a Galactic extinction \citep{Schlafly2011} toward NGC 384 of $A_H = 0.032$ mag.
  The projected dispersion along the major axis and the axis ratio for each component are given in Columns (3) and (4), respectively.
  Projected quantities are listed with primes.
  All components have a PA of $136.212^\circ$ east of north.}
\end{singlespace}
\label{tab_mgengcpgc}
\end{deluxetable}

\subsection{ALMA Observations\label{alma}}

We obtained ALMA band 6 observations of NGC 384 on 2016 October 13 through Program 2016.1.01010.S (PI: Walsh) in Cycle 4.
The observations consisted of a single pointing with one spectral window centered at 227.325 GHz, the redshifted frequency of the 230.538 GHz $^{12}$CO(2$-$1) emission line, along with two spectral windows centered on continuum with average frequencies of 229.288 GHz and 243.015 GHz.
The observations were taken in the C40$-$6 configuration with minimum (maximum) baselines of 16.7 m (3100 m) and a total on-source exposure time of 27.3 minutes.
Although dust is detected in the spectral windows centered on continuum, this work focuses on the CO emission.

We processed the data with Common Astronomy Software Applications (CASA) version 4.7.2, employing a \texttt{TCLEAN} deconvolution with Briggs weighting ($r = 0.5$; \citealt{Briggs1995}).
We used the line-free channels to perform $uv$-plane continuum subtraction.
The resultant synthesized beam full width at half maximum (FWHM) is $0\farcs{30}$ ($0\farcs{16}$) along the major (minor) axis, for a geometric mean of $0\farcs{22}$, or 59.0 pc.
The beam PA is $16.17^\circ$ east of north.
Fluxes were calibrated using the ALMA standard quasar J$2253+1608$ for flux calibration, leading to a $10\%$ uncertainty in the absolute flux calibration at this frequency \citep{Fomalont2014}.

Ultimately, the data cube has 124 frequency channels with widths of 15.07 MHz, corresponding to $\sim$19.88 \kms\ at the redshifted CO($2-1$) frequency.
We detect CO emission in channels 43 through 78, corresponding to recessional velocities of $cz = 3880.0 - 4575.7$ \kms.
The data cube's pixel scale is $0\farcs{04}$ pixel$^{-1}$.
Emission-free regions have root-mean-square (rms) noise at the 0.3 mJy beam$^{-1}$ channel$^{-1}$ level.

\subsubsection{Properties of the CO($2-1$) Emission\label{emission}}

\begin{figure*}
\includegraphics[width=\textwidth]{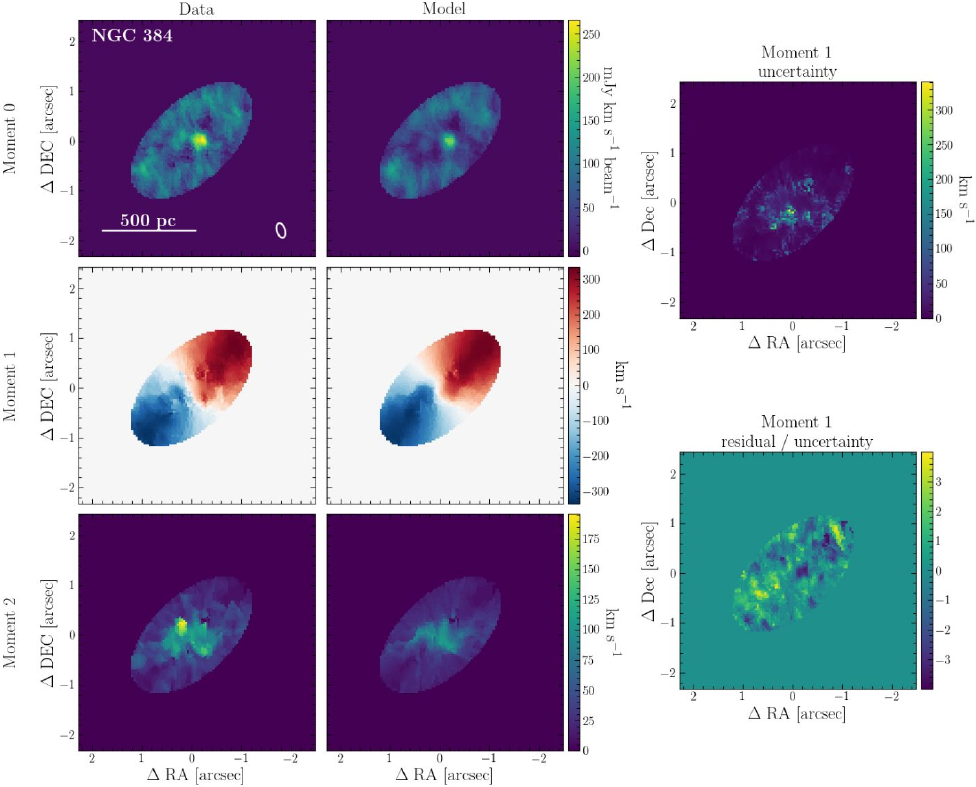}
\caption{Zeroth (\textit{top row}), first (\textit{middle row}), and second (\textit{bottom row}) moment maps of NGC 384 built from the ALMA data (\textit{left column}) and best-fit fiducial model (\textit{middle column}) within the fiducial elliptical fitting region.
The uncertainty in the first moment map (\textit{upper right}) and first moment map residual (data$-$model) normalized by the uncertainty (\textit{lower right}) are also displayed.
The moment maps are constructed on the original ALMA pixel scale of $0\farcs{04}$ pixel$^{-1}$ and linearly mapped to their respective scale bars, with each moment's data and model using the same scale.
These maps are not used in the fit, as models are fit directly to the data cube (\S\ref{model}).
The first moment map shows that the CO disk displays a kinematic twist in the inner $\sim0\farcs{5}$.
}
\label{fig_fiducial_moments_ngc}
\end{figure*}

Spatially resolved zeroth (integrated CO($2-1$) emission), first (projected line-of-sight velocity, $v_{\mathrm{los}}$), and second (projected line-of-sight velocity dispersion, $\sigma_{\mathrm{los}}$) moment maps of the ALMA observations of NGC 384 are displayed in Figure \ref{fig_fiducial_moments_ngc}.
During construction, we interactively mask pixels without discernible CO emission, and the final maps are shown within the elliptical fitting region used in dynamical modeling (chosen to encompass almost all the emission in the disk, while minimizing the inclusion of noise; see \S \ref{model}).
To calculate the uncertainty in the first moment, we perform a 1000-iteration Monte Carlo simulation.
In each iteration, we perturb the data cube, drawing each pixel from a Gaussian centered on the observed pixel value with a width given by the rms noise calculated in the given channel.
Then, we construct the first moment map from that new data cube.
The standard deviation of the resultant 1000 moment maps is taken as the moment map uncertainty in each pixel.

The CO emission is co-spatial with the NGC 384 dust disk, as seen in Figure \ref{fig_hst_ngc}.
The emission also displays a bright knot near the center and a dearth of flux both northwest and southeast of the knot.
The first moment map shows that the northwest side of the disk is redshifted and the southeast side is blueshifted, with line-of-sight velocities peaking at $\sim\pm350$ \kms. 
There is also a kinematic twist in the first moment map, which we account for in the modeling in \S\ref{warpeddisk}.
The second moment map peaks at $201$ \kms, northeast of the disk center.
At the disk center, the velocity dispersion reaches $\sim$120 \kms, and it drops to $\sim$20 \kms\ at a projected radius of $\sim$1\farcs5.

\begin{figure}[h!]
\includegraphics[width=0.47\textwidth]{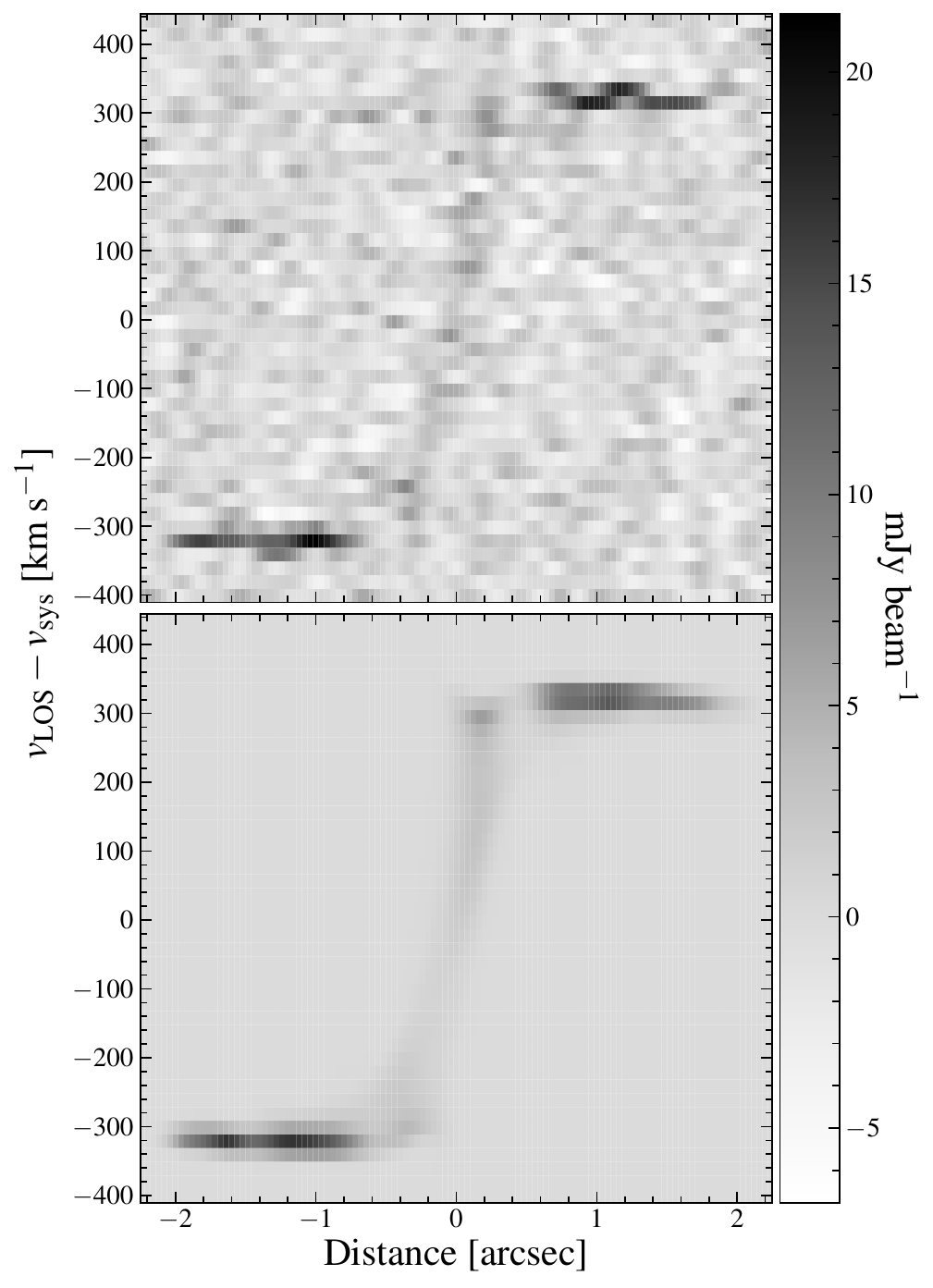}
\caption{PVDs extracted from the observed data cube (top) and best-fit fiducial model cube (bottom) for NGC 384.
The best-fit systemic velocity (see Table \ref{tab_fiducial_ngcpgc}) has been subtracted from the line-of-sight velocities displayed here.
The extraction path for the PVDs is along the disk major axis (PA $=132.8^\circ$ east of north, based on the flat disk model in \S\ref{error_ngc}), with a width set by the geometric mean of the synthesized beam ($0\farcs{22}$).
Both panels are linearly mapped according to the color bar.
As a result of the kinematic twist, the central region of the PVD highlights features of the minor kinematic axis.
Our dynamical models are fit directly to the data cube, so these PVDs are not used in the fit.
}
\label{fig_pvds}
\end{figure}

We extract flux densities along the major axis of the data cube with an extraction width equal to the geometric mean of the ALMA beam to construct position-velocity diagrams (PVDs).
For the major axis, we use an angle of $132.8^\circ$, measured east of north to the blueshifted side of the disk, corresponding to the best-fit flat disk model disk PA (see \S\ref{error_ngc}).
Due to the kinematic twist near the nucleus, the central region of the PVD primarily contains features that are along the minor kinematic axis.
The observed PVD is shown in Figure \ref{fig_pvds}.

The total CO(2$-$1) flux is $7.20 \pm 0.06\ \mathrm{(stat)} \pm 0.72\ \mathrm{(sys)}$ Jy km s$^{-1}$, with systematic uncertainties stemming from the flux calibration.
We estimate the CO(2$-$1) luminosity, $L^\prime_{\mathrm{CO(2-1)}}$, from the observed flux following \citet{Carilli2013}.
In order to convert this value to a CO(1$-$0) luminosity ($L^\prime_{\mathrm{CO(1-0)}}$), we assume $R_{21} \equiv L^\prime_{\mathrm{CO}(2-1)}/L^\prime_{\mathrm{CO}(1-0)} = 0.7$ \citep{Lavezzi1999}.
We then convert this luminosity to an H$_2$ mass, using the conversion factor $\alpha_\mathrm{CO}\equiv M_{\mathrm{H}_2}/L'_\mathrm{CO(1-0)} = 3.1\ M_\odot$ pc$^{-2}$ (K km s$^{-1}$)$^{-1}$ \citep{Sandstrom2013}, and a total gas mass using the helium-to-hydrogen mass ratio $f_\mathrm{HE} = 0.36$, such that $M_\mathrm{gas} = M_{\mathrm{H}_2} (1+f_{\text{HE}})$.
We find a total gas mass of ($7.91\pm0.07$ [stat] $\pm0.79$ [sys]) $\times10^7\ M_\odot$, which is on the same order of magnitude as the gas masses found in other ETGs with CO emission (e.g., \citealt{Boizelle2017,Ruffa2019,Ruffa2023}).
We likely underestimate the total systematic uncertainties in the gas mass, as we used an $\alpha_\mathrm{CO}$ calibrated for spiral galaxy disks, which may be different from ETG disks.

\section{\label{model}Dynamical Modeling}

We use a molecular gas-dynamical modeling code developed and tested in previous work \citep{Cohn2021,Cohn2023}.\footnote{Although this code is not currently published, anyone who wishes to reproduce our results may contact the corresponding author for the code.}
Following the example of our past work (e.g., \citealt{Barth2016b,Boizelle2019,Cohn2021,Cohn2023,Kabasares2022}), we assume the molecular gas follows circular orbits in a thin disk.
We calculate the circular velocity ($v_c$) relative to the systematic velocity ($v_\mathrm{sys}$) as a function of disk radius, based on the enclosed BH and stellar mass.
We treat the gas mass as negligible, although this assumption is tested in \S\ref{error_ngc}.
To calculate the enclosed stellar mass, we multiply the stellar light distribution by the stellar mass-to-light ratio (\ml) and deproject the dust-masked MGE, assuming an oblate axisymmetric shape with inclination angle (\inc) matching that of the flat gas disk.
We generate the circular velocities on a grid oversampled relative to the ALMA data cube by a factor of $s = $ \oversnp\ and converted to $v_\mathrm{los}$ using the gas disk \inc\ and position angle (\pa).

We build intrinsic line profiles along the observed ALMA data cube's frequency axis, with a channel spacing of 15.07 MHz.
The line profiles are assumed to be Gaussian, centered on $v_\mathrm{los}$ at each subsampled point, with a width ($\sigma_\mathrm{turb}$) that we take to be constant (\sig) throughout the disk.
The line profiles are weighted by the intrinsic CO flux map, which we estimate with a 10-iteration deconvolution of the zeroth moment map from the ALMA beam using the \texttt{lucy} task \citep{Richardson1972,Lucy1974} from \texttt{scikit-image} \citep{Walt2014}.
The flux in each pixel is divided evenly among the $s \times s$ subsampled pixels.
A scale factor of order unity, \fw, accounts for normalization mismatches between the observed and modeled line profiles.
Model line profiles are then summed back to the original pixel scale of the ALMA data cube, and each frequency slice of the model is convolved with the ALMA synthesized beam.
The model and data cubes are then down-sampled in bins of $\ds\times\ds$ spatial pixels to mitigate correlated noise \citep{Barth2016b}.

Finally, the model and data are compared directly within an elliptical fitting region that includes nearly all of the CO emission in each channel while omitting excess noise.
The fitting region contains 43 velocity channels, corresponding to $|v_\mathrm{los}-v_\mathrm{sys}|\lesssim 430$ km s$^{-1}$, with a projected semi-major axis of $\mathcal{R}_\mathrm{fit} = 1\farcs{5}$, an axis ratio $q_\mathrm{ell} = 0.54$, and a position angle \pa$_\mathrm{ell} = 132.8^\circ$ east of north.
Ultimately, in the fiducial model (see \S\ref{expdisk} and \S\ref{results}), the fitting region contains 6493 data points with 6481 degrees of freedom.\footnote{Fully accounting for the non-independence of adjacent $4 \times 4$ pixel groups that are separated by less than a full synthesized beam (see, e.g., Section \ref{error_ngc}) would translate to fewer degrees of freedom and higher estimates of $\chi^2_\nu$.
Such a change would not affect which dynamical model is judged to be best-fitting or alter the reality that systematic uncertainties dominate over statistical uncertainties in this case (Section \ref{error_ngc}).
We therefore defer consideration of this subtlety to a future paper.}

We adopt a likelihood of $L \propto \exp(-\chi^2/2)$, taking $\chi^2 = \sum_{j} ((d_j - m_j)^2/\sigma_j^2)$,
where $d_j$ are the down-sampled data points, $m_j$ the down-sampled model points, and $\sigma_j$ the noise in channel $j$, calculated as the standard deviation of an emission-free area of the down-sampled data.
We optimize the free parameters [\mbh, \ml, \inc, \pa, \vsys, \sig, the BH location (\xloc, \yloc), and \fw] with {\tt dynesty} \citep{Speagle2020}, a nested sampling code.
Flat priors are sampled for each parameter uniformly in linear space, except for \mbh, which is sampled uniformly in logarithmic space.
The $68\%$ and $99.7\%$ confidence intervals of the parameter posterior distributions are reported as $1\sigma$ and $3\sigma$ uncertainties.
See \citet{Cohn2021} for more model details.

\subsection{\label{warpeddisk}Warped Disk}
As shown in the first moment map in Figure \ref{fig_fiducial_moments_ngc}, the inner $\sim$0\farcs{5} of the disk appear to display a twist.
We find that a flat disk model must be oversimplified, as it is incapable of reproducing this kinematic twist.
Therefore, we implement multiple methods to account for the twist, allowing \inc\ and \pa\ to vary with radius.
The resultant model retains circular orbits for the gas, but adjacent orbits are no longer required to exist in the same plane, thus creating a warped disk.

\subsubsection{\label{tiltedrings}Tilted Ring Model}
We first implement a set of variable inclinations and position angles to model the warped disk, preserving the thin disk assumption, following a similar approach to the concentric tilted ring model developed in \citet{Boizelle2019}.
Here, we choose an integer number $N$ of radial nodes within the fitting region at which \pa$_n$ and \inc$_n$ are free parameters.
We create continuous \pa($r$) and \inc($r$) profiles by interpolating \pa\ and \inc\ linearly between these nodes, then extending \inc($r$) and \pa($r$) from the innermost node to $r=0$, and from the outermost node to $\mathcal{R}_\mathrm{fit}$.
In order to calculate the initial disk radii $r$, initial \pa\ and \inc\ are required, so we use the best-fit \pa\ and \inc\ from a flat disk model (see \S\ref{error_ngc}).
A new warped disk radius $R$ grid is then calculated from \inc($r$) and \pa($r$), and the rest of the model is generated as described in \S\ref{model} on this grid.
This approach avoids potential interpolation concerns when applying tilted-ring models to more moderately warped disks.

Given the relatively low signal-to-noise ratio of the NGC 384 observations compared to the data in \citet{Boizelle2019}, we test the simplest case, using two nodes evenly spaced over the fiducial fitting ellipse.
With a free \inc\ and \pa\ at each of these two nodes, this model includes four new free parameters, which are fit simultaneously with the other model parameters.
We test the exact choice of node location in \S\ref{error_ngc} below, finding that results for the model with two nodes are robust and consistent, although the tilted ring \pa\ is incapable of replicating the central twist in the data.

\subsubsection{\label{expdisk}Parameterized Warped Disk}
The tilted ring model constructed for NGC 384 is stable in the case with two nodes, but a tilted ring model with only two nodes is incapable of producing the rapid \pa\ warp seen at the center of the first moment map (see Figure \ref{fig_fiducial_moments_ngc}).
We attempt to construct tilted ring models with $N\geq3$ to account for this warp, but due to the low signal-to-noise of the data and the increasing number of free parameters, these models do not exhibit consistent behavior or provide robust parameter solutions.
As such, we develop a new warped disk model using a functional form to allow greater radial changes in the \pa\ and \inc\ parameters, without drastically increasing the number of free parameters in the model.

First, we test a model allowing \inc\ and \pa\ to vary linearly across the entire disk, which we refer to as the ``linear twist'' model.
In this model, there are two free \inc\ and two free \pa\ parameters, one each at the disk center and one each at the disk edge.
As with the tilted ring model, calculating the disk radii $r$ requires an initial \inc\ and \pa, for which we use the best-fit flat disk model's \inc\ and \pa.
The resultant linear \inc$(r)$ and \pa$(r)$ functions are used to construct a new grid of disk radii $R$, on which we build the rest of the model, as detailed in \S\ref{model}.
As with the tilted ring model above, this model includes four new free parameters.
As discussed in \S\ref{error_ngc}, \inc\ is well-described with a linear fit, but we find that a linear \pa\ is unable to account for the relatively flat \pa\ at large radii combined with the strong central twist.

As such, we adopt an exponential function, $\pa(r) = \pa_0+\pa_1\exp{(-r/r_\pa)}$, to allow \pa\ to warp more strongly at the center.
However, we continue to keep the inclination linear in this model, with \inc$_0$ free at the disk center and \inc$_1$ free at the disk edge.
There are five free parameters in this model: $\inc_0, \inc_1, \pa_0, \pa_1$, and $r_\pa$.
Comparison of this parameterized warped disk modeling method showed very similar results for NGC 384 to the tilted ring modeling method presented by \citet{Boizelle2019}, with agreement typically within a couple \kms\ and only reaching 10$-$20 \kms\ in the innermost couple beam areas.
Finally, we also test a parameterized warped disk model that uses an exponential function for both inclination and position angle.
This more complex model returns an \inc$(r)$ that is consistent with a linear function, so we move forward with the parameterization using an exponential \pa$(r)$ and linear \inc$(r)$ in our fiducial model.

\section{\label{results}Modeling Results}
Below, we discuss the results of our dynamical modeling of the disk in NGC 384.
The fiducial warped disk model, with \inc\ and \pa\ parameterized as linear and exponential functions, respectively, captures the kinematic twist seen in the moment map and yields a significantly improved $\chi^2$ and reduced $\chi^2$ ($\chi^2_\nu$) over the flat disk, tilted ring, and linear twist models.

\subsection{\label{n384results}Fiducial Model Results}

\begin{figure*}
\includegraphics[width=\textwidth]{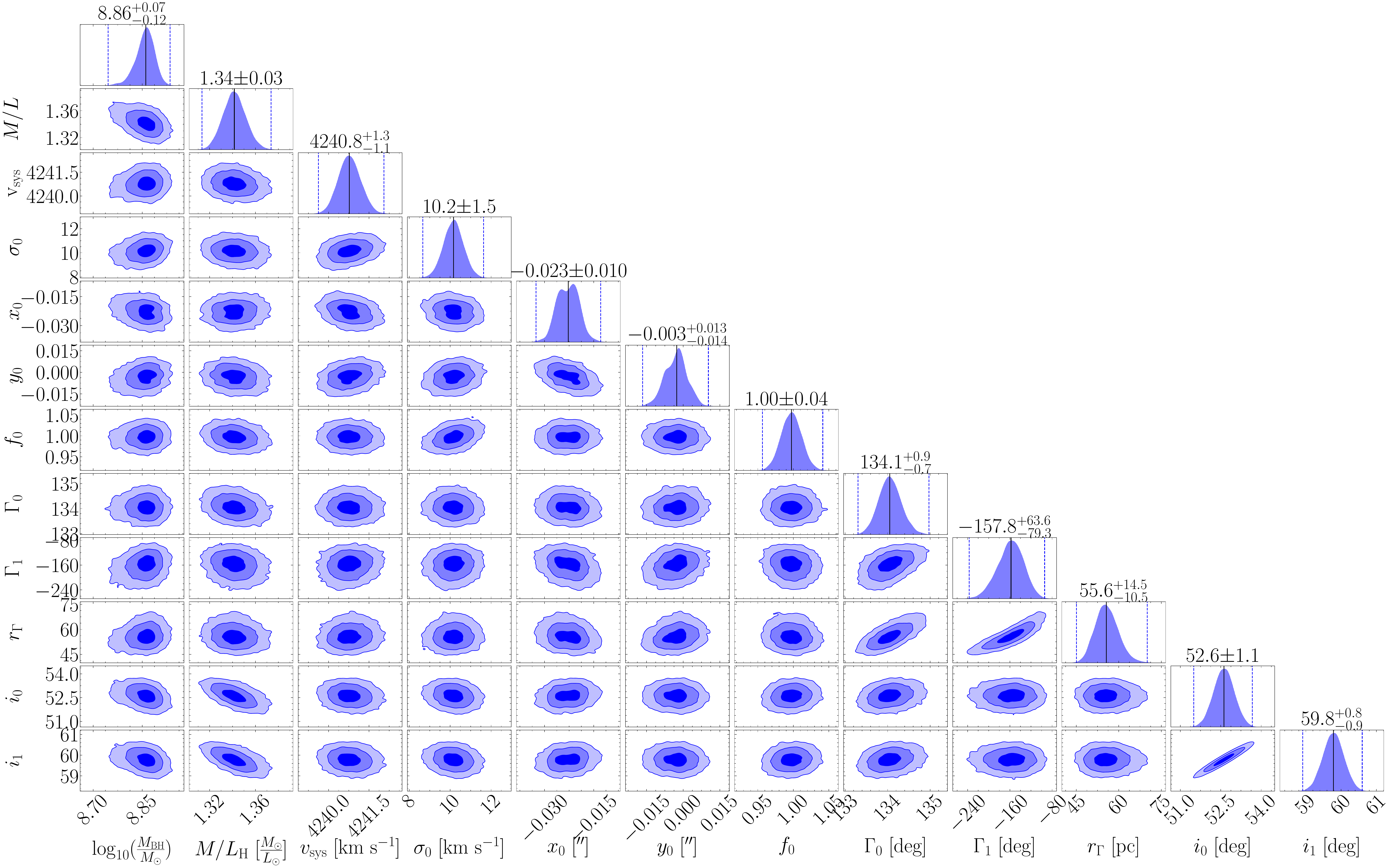}
\caption{One-dimensional (1D; top edge) and 2D posteriors of fiducial model parameters for NGC 384.
The 2D panels display $1\sigma$, $2\sigma$, and $3\sigma$ contours.
In 1D panels, black lines correspond to the posterior median and blue dashed lines to $3\sigma$ confidence intervals.
Parameter median and $3\sigma$ values are also labeled above each 1D panel.
The BH mass is \mbh\ $= (\mbhmed^{+\mbhups}_{-\mbhdowns} [1\sigma] ^{+\mbhupsss}_{-\mbhdownsss} [3\sigma]) \times 10^8\ M_\odot$.
Our best-fit fiducial model is built with the posterior medians and produces a reduced $\chi^2_\nu = 1.159$.
}
\label{fig_fiducial_corner_ngc}
\end{figure*}

The best-fit parameters for the fiducial model of NGC 384 are listed in Table \ref{tab_fiducial_ngcpgc} and posterior distributions of the free parameters are displayed in Figure \ref{fig_fiducial_corner_ngc}.
From the best-fit model, we construct moment maps and PVDs, which are shown on the original ALMA $0\farcs{04}$ pixel$^{-1}$ scale in Figures \ref{fig_fiducial_moments_ngc} and \ref{fig_pvds}, respectively.
The uncertainty and residual maps of the first moment are also also shown in Figure \ref{fig_fiducial_moments_ngc}.
In Appendix \ref{appendix}, we compare the fiducial model and observed line profiles on the down-sampled pixel scale over the full fitting ellipse.
The parameterized warped disk \inc\ and \pa\ profiles are shown in Figure \ref{fig_warped_profiles}, reflecting strong constraints on a linear twist in the inclination and a significant central twist in the position angle.

We find \mbh\ = $(\mbhmed^{+\mbhups}_{-\mbhdowns}[^{+\mbhupsss}_{-\mbhdownsss}]) \times 10^8\ M_\odot$ ($1\sigma$ [$3\sigma$] uncertainties), with $\chi^2 = 7511.1$ and $\chi^2_\nu = 1.159$.
We also determine \ml\ $ = \mlmed\pm\mldowns[\pm\mldownsss]\ M_\odot/L_\odot$.
Using a $\sim12-13.5$ Gyr stellar age and a metallicity $\sim$0.05 dex above solar, which match the age and metallicity derived for NGC 384 in \cite{Yildirim2017}, simple stellar population models \citep{Vazdekis2010} suggest a similar, albeit slightly lower, \ml\ $\sim 1.2$ $M_\odot/L_\odot$ for a \cite{Kroupa2001} initial mass function (IMF).
Our results are thus consistent with more bottom-heavy IMFs seen at the centers of massive ETGs (e.g., \citealt{Martin2015a,Martin2015,LaBarbera2019,Mehrgan2024}).

\begin{deluxetable}{llllc}[t]
\tabletypesize{\small}
\tablecaption{Modeling Results}
\tablewidth{0pt}
\tablehead{
\colhead{Parameter} & 
\colhead{Median} & 
\colhead{$1\sigma$} &
\colhead{$3\sigma$} &
\colhead{Prior range}
\\[-1.5ex]
\colhead{(1)} & 
\colhead{(2)} & 
\colhead{(3)} &
\colhead{(4)} &
\colhead{(5)}
}
\startdata
\mbh\ [$10^8\ M_\odot$] & ${\mbhmed}$ & $_{-\mbhdowns}^{+\mbhups}$ & $_{-\mbhdownsss}^{+\mbhupsss}$ & ${0.32} \rightarrow {10.00}$ \\
\ml\ [$M_\odot/L_\odot$] & ${\mlmed}$ & $^{+\mlups}_{-\mldowns}$ & $^{+\mlupsss}_{-\mldownsss}$ & ${0.50} \rightarrow {3.00}$ \\
$\inc_0$ [$^\circ$] & ${52.6}$ & $\pm0.4$ & $_{-1.1}^{+1.0}$ & ${0.0} \rightarrow {89.9}$ \\
$\inc_1$ [$^\circ$] & ${59.8}$ & $\pm0.3$ & $_{-0.9}^{+0.8}$ & ${0.0} \rightarrow {89.9}$ \\
$\pa_0$ [$^\circ$] & ${134.1}$ & $\pm0.3$ & $_{-0.8}^{+0.9}$ & ${120.0} \rightarrow {150.0}$ \\
$\pa_1$ [$^\circ$] & ${-157.8}$ & $_{-26.2}^{+23.5}$ & $_{-80.5}^{+64.8}$ & ${-270.0} \rightarrow {0.0}$ \\
$r_\pa$ [pc] & ${55.6}$ & $_{-3.8}^{+4.2}$ & $_{-10.6}^{+14.5}$ & ${0.0} \rightarrow {5000.0}$ \\
\vsys\ [km s$^{-1}$] & ${4240.8}$ & $\pm0.4$ & $_{-1.2}^{+1.3}$ & ${4200.0} \rightarrow {4300.0}$ \\
\sig\ [km s$^{-1}$] & ${10.2}$ & $\pm0.5$ & $\pm1.5$ & ${0.0} \rightarrow {30.0}$ \\
\xloc\ [\arcsec] & $-0.023$ & $\pm0.003$ & $\pm0.010$ & $-0.187 \rightarrow 0.133$ \\
\yloc\ [\arcsec] & $-0.003$ & $_{-0.004}^{+0.005}$ & $_{-0.014}^{+0.013}$ & $-0.168 \rightarrow 0.152$ \\
\fw\ & ${1.00}$ & $\pm0.01$ & $\pm0.04$ & ${0.50} \rightarrow {1.50}$
\enddata
\begin{singlespace}
\tablecomments{
Best-fit fiducial model results.
Free parameters are listed in column (1).
Median values of each parameter's posterior distribution are shown in column (2).
Statistical $1\sigma$ and $3\sigma$ uncertainties are given in columns (3) and (4), respectively.
Prior ranges are shown in column (5).
The position angle $\pa_0$ is measured east of north to the major axis on the blueshifted side of the disk, while the exponential coefficient $\pa_1$ is measured such that positive values represent a counterclockwise shift.
The (\xloc, \yloc) coordinates of the BH are given relative to RA = $01^{\mathrm{h}} 07^{\mathrm{m}} 25.0181^{\mathrm{s}}$ and Dec = $+32^\circ 17\arcmin 33\farcs{798}$ (J2000), the maximum of the continuum emission.
Positive \xloc\ values correspond to shifts eastward and positive \yloc\ values to shifts northward.
}
\end{singlespace}
\label{tab_fiducial_ngcpgc}
\end{deluxetable}

\subsection{Systematic Uncertainties \label{error_ngc}}

To accurately characterize the uncertainty in the BH mass, we must also account for the choices we made when constructing our models and how they affect \mbh\ (e.g., \citealt{Boizelle2019,Cohn2021,Cohn2023,Kabasares2022}).
Below, we describe the impact of various modeling assumptions on \mbh\ and \ml\ and characterize the systematic uncertainties implied by their departures from the fiducial model.
The overall systematic uncertainties on \mbh\ and \ml\ include terms (summed in quadrature) for all alternative models that we judge to provide acceptable fits to the data.
To decide whether a given model is ``acceptable'' in this context, we primarily consider the Bayesian Information Criterion (BIC) $= k\log(N) - 2\log(L)$, where $k =$ the number of free parameters, $N =$ the number of data points, and $L =$ the maximum likelihood.
We exclude models that have $\Delta$BIC $\geq+10$ (where $\Delta$BIC $=$ test model BIC $-$ fiducial model BIC), indicating the fiducial model is significantly favored \citep{Liddle2007}.
Therefore, the models that we include in our systematic uncertainty calculations are reasonable models that sufficiently reproduce the data.
For example, we test a model with \mbh\ fixed to 0, which produces a $7\%$ increase to \ml\ from the fiducial model.
However, this model is ruled out with $\Delta$BIC $=+66.4$ compared to the fiducial model, so we exclude it from our systematic uncertainty calculations.

\begin{figure}
\includegraphics[width=0.47\textwidth]{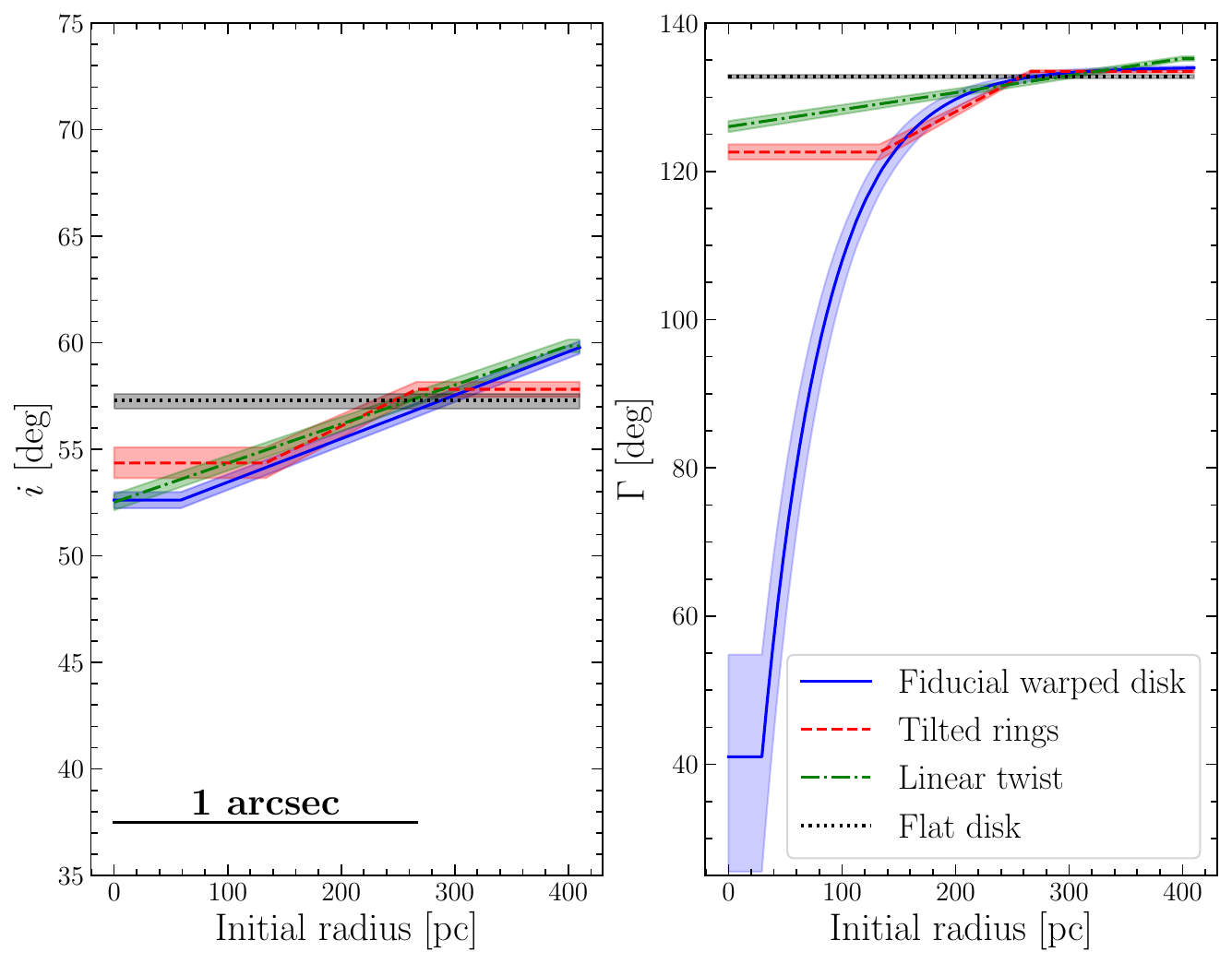}
\caption{Inclination (\inc; left) and position angle (\pa; right) profiles for the best-fit fiducial warped disk (blue solid line), tilted ring (red dashed line), linear twist (green dash-dotted line), and flat disk (black dotted line) models, plotted as a function of the input radius $r$ over the extent of the disk.
Shaded regions correspond to 1$\sigma$ confidence intervals.
The $\inc(r)$ for the fiducial warped disk model is consistent with the $\inc(r)$ profiles of the tilted ring and the linear twist models, while $\pa(r)$ for the fiducial warped disk model shows a significant, well-constrained twist at the center.
At radii $\gtrsim$200 pc ($\gtrsim$0\farcs{75}, outside of the central twist), the fiducial warped disk $\pa(r)$ agrees with the flat disk model.
}
\label{fig_warped_profiles}
\end{figure}

\textit{Warped disk.}
As discussed in \S\ref{warpeddisk}, in addition to the fiducial warped disk model, we test a tilted ring model to account for the kinematic twist in the CO disk.
Using the tilted ring model with two \inc\ and \pa\ nodes evenly spaced over the fitting region at $r=0\farcs{5}$ and $r=1\arcsec$, we find $\inc_0=54.4^\circ$, $\inc_1=57.8^\circ$, $\pa_0=122.6^\circ$, and $\pa_1=133.5$.
As shown in Figure \ref{fig_warped_profiles}, this \inc\ profile is consistent with the fiducial warped disk model.
The tilted ring \pa($r$) is also consistent with the fiducial model for $r\geq0\farcs{5}$, disagreeing only at the innermost radii where the tilted ring \pa\ is held constant.
In this model, \mbh\ decreases 47.0\% and \ml\ increases 3.9\% from the fiducial values, with an increased $\chi^2_\nu$ of 1.178.
We test a variety of node locations for the rings, ranging the inner node from 0\farcs{15}$-$0\farcs{6} and the outer node from 0\farcs{9}$-$1\arcsec.
In all cases, the model produces \inc\ and \pa\ profiles consistent with the tilted ring model with evenly spaced nodes.

Additionally, we consider a linear twist model, the warped disk model in which both \inc\ and \pa\ are linearly interpolated from the disk center all the way to the disk edge.
The resultant \inc\ profile is consistent with the fiducial warped disk model, while the \pa\ profile is relatively flat, dominated by the \pa\ at larger radii and failing to account for the changing \pa\ at the center (see Figure \ref{fig_warped_profiles}).
The \mbh\ in this model increases 20.3\% and \ml\ decreases 1.8\% from the fiducial model, and $\chi^2_\nu$ increases to 1.169.

We also test a flat disk model, with a single \inc\ and \pa\ value across the full disk, ignoring the observed kinematic twist.
This flat disk model has $\inc=57.3^\circ$ and $\pa=132.8^\circ$, consistent with the fiducial warped disk model results for radii $\gtrsim$200 pc ($\gtrsim$0\farcs{75}).
The flat disk model yields a 73.2\% lower \mbh\ relative to the fiducial model, a 6.3\% increase in \ml, and a much worse $\chi^2_\nu=1.195$.

The fiducial model is strongly preferred over all of the above models, with $\Delta\mathrm{BIC}=+112$ for the tilted ring model, $\Delta\mathrm{BIC}=+53$ for the linear twist model, and $\Delta\mathrm{BIC}=+211$ for the flat disk model.
Thus, we exclude all of these other disk structure models from our systematic uncertainty calculations as they are very strongly disfavored.

\textit{Dust extinction.}
We adopt the dust-masked MGE from Section \ref{mge} when fitting the fiducial dynamical model.
Here, we test the use of the original MGE, which ignores the presence of dust, and the dust-corrected MGE, which assumes $A_H=0.4$ mag to the southwest of the nucleus.
The dust-corrected MGE produces a positive shift to the inferred BH mass, with \mbh\ $ = 7.33\times10^8\ M_\odot$ (1.0\% larger than the fiducial \mbh), and a consistent \ml\ of 1.34 $M_\odot/L_\odot$.
This model has $\Delta$BIC $=-7.2$ compared to the fiducial model, indicating a better fit to the data.
The original MGE produces a negative shift to \mbh, with \mbh\ $ = 7.04\times10^8\ M_\odot$ (3.0\% lower than the fiducial \mbh) and \ml$=$1.37 $M_\odot/L_\odot$.
However, the original MGE model is a significantly worse fit than the fiducial model, with $\Delta$BIC $=+44.2$, so we exclude it from the systematic uncertainty calculations.

\textit{Radial motion.}
The kinematic twist near the center of the first moment map in NGC 384 could also reflect radial flows of the molecular gas.
Following \citet{Cohn2021} and \citet{Cohn2023}, we employ two toy models to determine whether the data favor any radial motion.
The first model introduces a radially constant velocity term ($v_{\mathrm{rad}}$), which is projected into the line of sight and summed with $v_{\mathrm{los}}$.
We find $v_{\mathrm{rad}}=13.3\pm8.4$ \kms\ ($3\sigma$ uncertainties), consistent with a low-velocity outflow.
\mbh\ decreases by 3.1\% and \ml\ decreases by 0.9\%.
The $v_{\mathrm{rad}}$ model produces the best fit of all of our models, with $\chi^2_\nu=1.156$, for a $\Delta$BIC compared to the fiducial model of $-13.2$.

In the second model, we use the dimensionless, radially varying term $\kappa$ \citep{Jeter2019}, which is multiplied by $v_c$, projected onto the line-of-sight, and summed with $v_{\mathrm{los}}$.
This model also favors a small outflow, with $\kappa=0.03\pm0.02$ (3$\sigma$ uncertainties).
From this best-fit $\kappa$, the median radial velocity in the disk is 9.0 \kms, consistent with the best-fit $v_{\mathrm{rad}}$ found above.
The best-fit \mbh\ decreases by 2.9\% from the fiducial model and \ml\ decreases by 0.9\%, with $\Delta$BIC compared to the fiducial model of $-12.0$.

\textit{Gas mass.}
Here, we test a model accounting for the molecular gas mass.
The radial CO surface brightness is measured within elliptical annuli on the zeroth moment map and used to calculate projected mass surface densities.
We then calculate circular velocities due to the gas mass ($v_{c,\mathrm{gas}}$) in the galaxy midplane by integrating following \citet{Binney2008}, assuming a thin disk.
The resultant $v_{c,\mathrm{gas}}$ is summed in quadrature with $v_c$ due to stars and the BH.
This model produces a 1.1\% increase in \mbh\ and 0.8\% decrease in \ml, fully consistent with the fiducial model's \mbh\ and \ml\ within uncertainties.
The maximum disk $v_{c,\mathrm{gas}}$ is $\sim$33 \kms, $\sim$8\% of the maximum circular velocity due to stars.
This model has $\chi^2_\nu=1.158$, with $\Delta$BIC $=-6.5$.

\textit{Turbulent velocity dispersion.}
In addition to the radially constant $\sigma_{\mathrm{turb}}$ used in our fiducial model, we test an exponential $\sigma_{\mathrm{turb}}(R) = \sigma_0 + \sigma_1 \exp(-R/R_\sigma)$ as a function of the warped disk radius.
This model converges to $\sigma_0 = 10.1$ \kms\ (consistent with the fiducial model), $R_\sigma = 22.6$ pc (0\farcs{085}), and $\sigma_1=221.4$ \kms, although the 3$\sigma$ confidence interval on $\sigma_1$ extends over the full 0 $-$ 500 \kms\ prior range.
In this model, \mbh\ decreases by 6.1\% and \ml\ increases by 0.4\%.
However, $\Delta$BIC $=+10.9$ compared to the fiducial model, indicating the exponential $\sigma_{\mathrm{turb}}$ is significantly disfavored, so we exclude it from the systematic uncertainty calculations.

\textit{Oversampling factor.}
It is possible for both ionized (e.g., \citealt{Barth2001}) and molecular (e.g., \citealt{Boizelle2019}) gas-dynamical \mbh\ measurements to depend on the pixel oversampling factor $s$.
In addition to the fiducial $s = \oversnp$, we test $s = $1, 2, 3, 4, and 8.
The greatest positive (negative) change to \mbh\ is $+0.5$\% ($-10.0\%$) with $s=8$ ($s=1$).
The greatest positive (negative) change to \ml\ is $+0.4$\% ($<-0.1$\%) with $s=1$ ($s=8$).
In all of these models, the $\Delta$BIC compared to the fiducial model is small, ranging from $-2.9$ ($s=2$) to $+2.1$ ($s=1$).

\textit{Intrinsic flux map.}
Here, we examine how the number of Lucy-Richardson deconvolution iterations on the CO flux map affect our results.
The fiducial model utilizes ten iterations, and we also test five and fifteen.
The five-iteration model produces a 0.3\% increase in \mbh\ and a 0.1\% increase in \ml.
The 15-iteration model produces a 1.2\% decrease in \mbh\ and an identical \ml.
The five-iteration model has $\Delta$BIC $=-5.4$ compared to the fiducial model and the fifteen-iteration model has $\Delta$BIC $=+9.1$, just within our threshold to include the model in the systematic uncertainty calculations.

\textit{Down-sampling factor.}
In the fiducial model, pixels are down-sampled in groups of $\ds\times\ds$ to mitigate correlated noise.
The synthesized ALMA beam size is 0\farcs{161} $\times$ 0\farcs{302} (4.0 $\times$ 7.6 pixels), and it has a position angle of $16.17^\circ$ east of north, such that the major axis of the beam is mostly aligned with the observed $y$-axis.
Therefore, we test a down-sampling factor of $4\times8$ spatial pixels.
We find that \mbh\ decreases by 5.0\% and \ml\ increases by 1.2\%.
However, the change in pixel binning means this test and the fiducial model have different data, and a direct BIC comparison between these models is not possible.
Nevertheless, this model is a significantly worse fit than the fiducial model, with $\chi^2_\nu=1.188$.
This is worse than the $\chi^2_\nu$ of almost every model that we have excluded by BIC comparisons, so we also exclude this model from the final systematic uncertainty calculations.

\textit{Fitting ellipse.}
Here, we vary the semi-major axis of the fitting ellipse, testing $\mathcal{R}_\mathrm{fit} = 1$\farcs{3} and 1\farcs{7}.
Fitting regions larger than 1\farcs{7} would include many noisy pixels outside of the CO disk.
The model with $\mathcal{R}_\mathrm{fit} = 1\farcs{3}$ produces a $6.0\%$ increase in \mbh\ and a 1.2\% increase in \ml.
In contrast, the $\mathcal{R}_\mathrm{fit} = 1\farcs{7}$ model finds \mbh\ increases by 4.5\% and \ml\ decreases by 1.0\%.
As changing $\mathcal{R}_\mathrm{fit}$ means including or excluding different data, direct BIC comparisons are again not possible.
Both models are somewhat worse fits than the fiducial model, with $\chi^2_\nu=1.163$ for $\mathcal{R}_\mathrm{fit} = 1\farcs{3}$, and $\chi^2_\nu=1.164$ for $\mathcal{R}_\mathrm{fit} = 1\farcs{7}$.
However, these $\chi^2_\nu$ values are comparable to those of models included by BIC comparisons, so we include them in our systematic uncertainty calculations.

\textit{Final error budget}.
We calculate the positive and negative systematic (sys) uncertainties on \mbh\ and \ml\ by summing the respective changes in quadrature from all of the tested models with acceptable BIC and $\chi^2_\nu$ values.
The greatest positive and negative shifts to \mbh\ are $+4.5$\% and $-10.0$\%, and come from the $\mathcal{R}_\mathrm{fit}=1\farcs{7}$ model and oversampling $s=1$ model, respectively.
The greatest positive and negative shifts to \ml\ are $+1.2$\% and $-1.5$\% from the $\mathcal{R}_\mathrm{fit}=1\farcs{3}$ and $1\farcs{7}$ models, respectively.
Thus, the BH mass is \mbhbothsigngc, and the \ml\ is \mlbothsig.

\section{Discussion \label{discussion}}

Our work, which reports the first dynamical \mbh\ measurement for NGC 384, means the number of molecular gas-dynamical \mbh\ determinations for local compact galaxies now equals the number of stellar-dynamical measurements in the sample.
We discuss the BH SOI in \S\ref{soi}, compare NGC 384 and the other relic galaxies with dynamical \mbh\ measurements to the local BH$-$host galaxy scaling relations in \S\ref{scaling_relations}, and discuss the significance of our results with respect to the co-evolution of BHs and galaxies in \S\ref{bh_gal_growth}.

\subsection{The BH Sphere of Influence and Comparisons to Literature\label{soi}}

The BH SOI, defined here as the radius at which the enclosed stellar mass is equal to \mbh, is $r_\mathrm{SOI} = 0\farcs{12}$ (31 pc) for our fiducial model.
If we instead estimate the BH SOI as $r_g = G$\mbh$/\sigma_\star^2$ and take $\sigma_\star = 221$ km s$^{-1}$ \citep{Yildirim2017}, we find $r_g = 0\farcs{24}$ (64 pc).
To quantify how well the data resolve the BH SOI, we compare the SOI to the geometric mean of the ALMA beam ($\theta_\mathrm{FWHM}=0\farcs{22}$) via $\xi = 2r_\mathrm{SOI}/\theta_\mathrm{FWHM}$ \citep{Rusli2013b}.
Using $r_\mathrm{SOI}$ and $r_g$, we find $\xi =$ 1.1 and 2.2, respectively.
These results are comparable to many other ALMA dynamical \mbh\ measurements, which often have $\xi\sim1-2$ (e.g., \citealt{Barth2016b,Onishi2017,Davis2017,Davis2018,Smith2019,Smith2021,Nguyen2020,Cohn2021,Cohn2023,Kabasares2022,Ruffa2023,KabasaresCohn2024}).

In NGC 384, the 1$\sigma$ statistical uncertainties in \mbh\ are at the $\sim$6$-$7\% level, while the statistical uncertainties in \mbh\ for UGC 2698 and PGC 11179 are at the 2$-$3\% level \citep{Cohn2021,Cohn2023}.
The total systematic uncertainties on \mbh\ in NGC 384 remain larger at the level of $\sim$8$-$14\%.
These results continue to demonstrate that accounting for the systematic uncertainties in molecular gas-dynamical modeling is critical for accurately characterizing the confidence in the measured \mbh.

\subsection{Comparing the Compact Relic Galaxies to the Local BH Scaling Relations \label{scaling_relations}}

\begin{figure*}
\centering
\includegraphics[width=\textwidth]{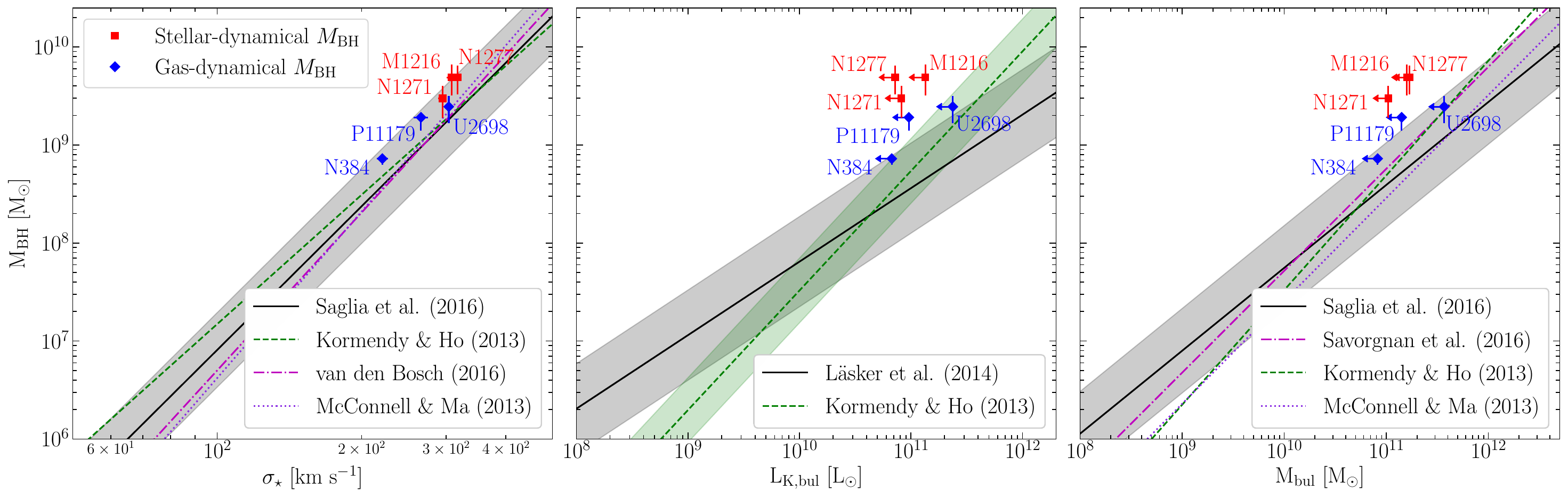}
\caption{The \msig\ (left), \mlum\ (middle), and \mmass\ (right) relationships \citep{Kormendy2013,McConnellMa2013,Lasker2014,Bosch2016,Saglia2016,Savorgnan2016}, with shaded regions indicating their intrinsic scatter.
The local compact relic galaxies with ALMA-based molecular gas-dynamical \mbh\ determinations \citep{Cohn2021,Cohn2023}, including NGC 384 (this work), are shown with blue diamonds.
Red squares indicate stellar-dynamical \mbh\ measurements from adaptive optics-assisted integral field spectroscopy \citep{Walsh2015,Walsh2016,Walsh2017}.
The relic galaxies are plotted with their total luminosities and masses on the \mlum\ and \mmass\ relations, making them upper limits on bulge values.
PGC 11179, Mrk 1216, NGC 1271, and NGC 1277 are positive outliers from \mlum\ and \mmass, but NGC 384 and UGC 2698 are consistent within the intrinsic scatter of all three relations.
}
\label{fig_scaling_relations}
\end{figure*}

In Figure \ref{fig_scaling_relations}, we compare our measurement of \mbh\ in NGC 384 to the BH scaling relations.
For the \msig\ relation, we use the stellar velocity dispersion of NGC 384 measured within a circular aperture at the galaxy half-light radius, $\sigma_\star = 221\pm6$ km s$^{-1}$ \citep{Yildirim2017}.
For the \mlum\ relation, we calculate the total $H$-band luminosity from the dust-masked MGE, finding $L_H = 6.13\times10^{10}\ L_\odot$, then convert to $L_K$ using an absolute $H$-band ($K$-band) Solar magnitude of 3.37 (3.27) mag \citep{Willmer2018} and $H-K=0.2$ mag from SSP models \citep{Vazdekis2010}.
We find $L_K = 6.72\times10^{10}\ L_\odot$.
To calculate the total stellar mass, we multiply $L_H$ by the best-fit fiducial \ml, finding $M_{\star} = 8.22\times10^{10}\ M_\odot$.
The other local compact galaxies with stellar-dynamical (NGC 1271, NGC 1277, and Mrk 1216; \citealt{Walsh2015,Walsh2016,Walsh2017}) and molecular gas-dynamical (UGC 2698 and PGC 11179; \citealt{Cohn2021,Cohn2023}) \mbh\ measurements are also shown in Figure \ref{fig_scaling_relations}, using the host properties listed in \citet{Cohn2021} and \citet{Cohn2023}.
Given debates on the bulge properties of the local compact galaxies \citep{Graham2016b,SavorgnanGraham2016}, we use the total galaxy luminosity and stellar mass as upper bounds on bulge values when displaying these objects in Figure \ref{fig_scaling_relations}.
Total uncertainties are calculated for all of the BH masses in the figure by summing the systematic and the 1$\sigma$ statistical uncertainties in quadrature.
Unlike the majority of the local compact galaxies, NGC 384 lies within the scatter of all three scaling relations.

Compared to the \mlum\ relation of \citet{Kormendy2013} and the \mmass\ and \msig\ relations of \citet{Saglia2016}, the \mbh\ we measure for NGC 384 is a factor of 2.2, 2.2, and 1.9$\times$ above the expected value, respectively.
This result places the measurement within the upper end of the scatter of each relation, in contrast to the five previously measured BH masses in the local compact galaxy sample, which lie an average of 6.3, 7.4, and 2.0$\times$ above the expected scaling relation values for the \mlum, \mmass, and \msig\ relations, respectively.

Next, we quantify how offset the BH masses in the local compact galaxy sample as a whole are from the local scaling relations using a Monte Carlo simulation.
For each of the six objects with \mbh\ measurements to date, we draw one value from a normal distribution centered on the scaling relation-predicted \mbh\ with a width equal to the relation intrinsic scatter, and one from a distribution centered on the measured \mbh\ with a width given by the total measurement uncertainties.
In each iteration, there are thus six scaling relation BH masses and six dynamical BH masses.
We repeat this process 10,000 times and find that the median \mbh\ from the six dynamical BH masses is above the median \mbh\ from the six scaling relation-predicted BH masses 100.0\%, 99.9\%, and 91.5\% of the time (over the 10,000 iterations), for the \mlum\ \citep{Kormendy2013}, \mmass, and \msig\ \citep{Saglia2016} relations, respectively.
Moreover, the median offset between the two distributions is $2.1\times10^9\ M_\odot$, $2.0\times10^9\ M_\odot$, and $1.2\times10^9\ M_\odot$ for the \mlum, \mmass, and \msig\ relations, respectively.

These results still hold when we consider only the three ALMA measurements, in which case the median of the dynamically measured distribution is above the scaling relation-predicted distribution for \mlum, \mmass, and \msig\ in 97.2\%, 95.2\%, and 88.2\% of the iterations, respectively.
We caution that these comparisons are preliminary, as only six out of the sample of fifteen local compact relic galaxies have dynamical \mbh\ measurements so far.

Comparing to the \citet{Zhu2021} BH mass$-$core mass scaling relation, and using the total stellar mass as the core mass, our \mbh\ for NGC 384 is a factor of 1.02 above the predicted \mbh, consistent with the relation.
Likewise, the other five dynamical \mbh\ measurements in the relic sample are consistent with the \citet{Zhu2021} relation within its scatter.
This relation is constructed for classical bulges and the cores of elliptical galaxies, which are likely descendants of red nugget galaxies at $z\sim2$.

\subsection{BH and Host Galaxy Co-Evolution \label{bh_gal_growth}}

NGC 384 is a member of a sample of compact galaxies that are local analogs and likely relics of $z\sim2$ red nuggets \citep{Yildirim2017}.
It is a fast rotator with a disky shape, uniform old ($\sim$10 Gyr) stellar population, super-solar stellar metallicity, and an elevated central surface density that falls off steeply at large radii \citep{Yildirim2017}.
NGC 384 also has a red population of globular clusters \citep{Kang2021} and a small effective radius (1.5 kpc; \citealt{Yildirim2017}) for its stellar mass ($8.22\times10^{10}\ M_\odot$), consistent with the mass-size relation at $z\sim2$ \citep{Wel2014}.
These properties, typical of the local compact galaxy sample, are also consistent with $z\sim2$ red nuggets and the cores of giant ellipticals (e.g., \citealt{Trujillo2014,Mateu2015,Mateu2017,Martin2015,Yildirim2017}), which tend to host the most massive BHs in the local Universe.

Furthermore, $z\sim2$ red nuggets are thought to seed the cores of massive local ellipticals whose growth through dry mergers increases bulge stellar mass and luminosity without significantly feeding the BH (e.g., \citealt{Naab2009,Oser2010,Dokkum2010}).
NGC 384 and the other local compact galaxies are thus likely passively evolved relics of $z\sim2$ red nuggets that failed to undergo such mergers \citep{Yildirim2017}.
Such relics of red nuggets have also been observed in cosmological simulations \citep{Wellons2016}.
Although some of the local compact galaxies are isolated, most are located in group or cluster environments \citep{Yildirim2017}, and NGC 384 is a member of Arp 331 (also known as the Pisces Cloud), a galaxy group in Pisces \citep{Vaucouleurs1976}.
Nevertheless, the local compact galaxies have regular isophotes and no evidence of tidal interactions, indicating they likely have not formed through the stripping of outer layers.

Finding over-massive BHs in most or all passively evolved relics of $z\sim2$ red nuggets would suggest that BHs in massive ETGs tend to finish growing prior to the stars in the galaxy outskirts.
Most red nugget galaxies would then have to undergo sufficiently many dry mergers between $z\sim2$ and the present day for their stellar masses and luminosities to catch up to the local \mlum\ and \mmass\ relations.
We note that the sample's closer agreement to \msig\ despite positive offsets from \mlum\ and \mmass\ is unsurprising due to the fact that the Faber-Jackson relation levels off for power-law ETGs \citep{Lauer2007}.
Moreover, \citet{Matt2023} showed there is a difference in the predicted number density of high-mass BHs (\mbh\ $>10^9M_\odot$) at $1\lesssim z\lesssim3$ based on whether they assume \mmass\ versus \msig.
The fact that the local red nugget relic sample is more closely aligned with \msig\ may indicate that the local \msig\ relation is a more accurate representation of high-mass BHs at $1\lesssim z\lesssim3$ than the local \mmass\ relation.

Although previous work found evidence for over-massive BHs in
relic galaxies (e.g., \citealt{Mateu2015,Walsh2015,Walsh2016,Walsh2017,Cohn2023}), the local compact galaxies NGC 384 and UGC 2698 challenge this interpretation due to their consistency with the local BH scaling relations.
However, UGC 2698 may be a less pristine relic \citep{Yildirim2017} and thus could represent an intermediate step between the $z\sim2$ and $z\sim0$ BH scaling relations \citep{Cohn2021}.
NGC 384, on the other hand, is a more typical local compact galaxy, showing no indication of any substantial growth since $z\sim2$ \citep{Yildirim2017}.
Therefore, our gas-dynamical result could suggest that not all BHs in $z\sim2$ red nuggets were over-massive compared to the local scaling relations.

In this case, the properties of the local compact galaxies could be explained by much greater intrinsic scatter at the high-mass end of the scaling relations than has previously been estimated.
However, the apparent scatter could also be inflated by systematic differences in stellar- and gas-dynamical measurement methods.
Such systematics remain poorly understood, as there are currently only a handful of objects of any kind with \mbh\ measurements from both stellar- and ALMA-based gas-dynamics.
Nevertheless, in three of those cases, the stellar-dynamical measurement is a factor of $\sim$2$\times$ larger than the gas-dynamical measurement \citep{Krajnovic2009,Rusli2011,Schulze2011,Rusli2013a,Barth2016a,Davis2017,Boizelle2019,Smith2019,Dominiak2024b,Waters2024}.
Stellar-dynamical \mbh\ measurements for NGC 384, UGC 2698, and PGC 11179 may result in higher \mbh\ values that are more significant outliers from the scaling relations.
We note that there are five galaxies with dust disks in the relic sample that are excellent targets for additional molecular gas-dynamical \mbh\ measurements but as-yet have no ALMA observations.
Directly cross-checking these ALMA-based measurements with stellar dynamics, as well as obtaining stellar- and molecular gas-dynamical \mbh\ measurements for additional local compact galaxies, is crucial for accurately characterizing the scatter of the scaling relations and determining what implications the sample has for BH$-$host galaxy co-evolution.

\section{Conclusions\label{conclusions}}

We have observed CO($2-1$) emission from the circumnuclear gas disk in the local compact relic galaxy NGC 384 at 0\farcs{22} resolution with ALMA.
We measure spatially resolved kinematics of the gas disk, identifying a significant kinematic twist in the first moment map.
Therefore, we fit a dynamical model accounting for this twist, such that the disk inclination varies linearly with radius and the position angle varies exponentially with radius.
We test a variety of additional dynamical models, finding \mbhfullerrngc\ and stellar \mlfullerr.
The BH SOI is resolved by the data ($\xi=2r_\mathrm{SOI}/\theta_\mathrm{FWHM}\sim1-2$), and we find that the total systematic uncertainties are a factor of $\sim$2$\times$ larger than the statistical uncertainties, indicating systematics are vital to consider for molecular gas-dynamical \mbh\ measurements.
Obtaining higher signal-to-noise CO imaging would enable more detailed gas-dynamical modeling, likely shrinking measurement uncertainties.
We find NGC 384 lies within the upper end of the scatter of all three BH$-$host galaxy scaling relations.

NGC 384 is a likely relic of a $z\sim2$ red nugget, with an evolutionary history distinct from those of typical local massive ETGs \citep{Yildirim2017}.
However, its location relative to the BH scaling relations is different from the over-massive BHs found in other local compact relic galaxies studied thus far \citep{Walsh2015,Walsh2016,Walsh2017}, including one (PGC 11179) measured with molecular gas \citep{Cohn2023}.
Unlike UGC 2698, which is also consistent with all three scaling relations within their scatter \citep{Cohn2021}, NGC 384 shows no evidence of any mergers or growth since $z\sim2$.
These properties call into question previous evolutionary interpretations of the over-massive BHs in local compact relic galaxies, which had suggested those BHs might have grown prior to the growth of stars in the galaxy outskirts.

Our result may instead be evidence that there is greater intrinsic scatter than previously thought, due to a diversity of growth histories in the high-mass end of the scaling relations.
Another alternative is that $z\sim2$ galaxies may follow a much steeper \mlum\ relation than locally.
However, the dearth of direct comparisons between molecular gas-dynamical \mbh\ measurements and stellar-dynamical determinations calls into question whether this scatter may result partly from a systematic offset between the two methods.
Obtaining stellar-dynamical \mbh\ measurements to compare to the molecular gas dynamics, as well as making \mbh\ measurements for the remainder of the local compact relic galaxies, is required to determine whether the sample truly contains over-massive BHs.

\section*{Acknowledgements}
J.H.C. and J.L.W. were supported in part by NSF grant AST-1814799 and AST-2206219.
This paper makes use of the following ALMA data: ADS/JAO.ALMA\#2016.1.01010.S.
ALMA is a partnership of ESO (representing its member states), NSF (USA) and NINS (Japan), together with NRC (Canada), MOST and ASIAA (Taiwan), and KASI (Republic of Korea), in cooperation with the Republic of Chile.
The Joint ALMA Observatory is operated by ESO, AUI/NRAO and NAOJ.
The National Radio Astronomy Observatory is a facility of the National Science Foundation operated
under cooperative agreement by Associated Universities, Inc.
Based on observations made with the NASA/ESA Hubble Space Telescope, obtained from the data archive at the Space Telescope Science Institute, which is operated by the Association of Universities for Research in Astronomy, Inc., under NASA contract NAS5-26555.
The observations are associated with program \#13050.
The HST data presented in this paper were obtained from the Mikulski Archive for Space Telescopes (MAST) at the Space Telescope Science Institute.
The specific observations analyzed can be accessed via \dataset[DOI: 10.17909/emkb-rz64]{https://doi.org/10.17909/emkb-rz64}.
Portions of this research were conducted with the advanced computing resources provided by Texas A\&M High Performance Research Computing.
This work used arXiv.org and NASA's Astrophysics Data System for bibliographic information.
A.J.Bak. acknowledges support from the Radcliffe Institute for Advanced Study at Harvard University.
L.C.H. was supported by the National Science Foundation of China (11991052, 12011540375, 12233001), the National Key R\&D Program of China (2022YFF0503401), and the China Manned Space Project (CMS-CSST-2021-A04, CMS-CSST-2021-A06).

J.H.C. would like to thank Silvana Delgado Andrade, Kate Pitchford, Taylor Hutchison, Ryan Hickox, Quinn Casey, and Emmanuel Durodola for their support and helpful discussions.
We thank the anonymous referee for their valuable comments, which improved our manuscript.

\facilities{ALMA, HST (WFC3)}

\software{CASA (v4.7.2; \citealt{McMullin2007,CASA2022}), \texttt{dynesty} (\citealt{Speagle2020}, \dataset[DOI]{https://doi.org/10.5281/zenodo.7995596}), GALFIT \citep{Peng2010}, kinemetry \citep{Krajnovic2006}, mgefit \citep{Cappellari2002}, Tiny Tim \citep{Krist2004}, AstroDrizzle \citep{Gonzaga2012,Hack2012}, scikit-image \citep{Walt2014}, ASTROPY \citep{Astropy2013,Astropy2018,Astropy2022}, MATPLOTLIB \citep{Hunter2007}, NUMPY \citep{Walt2011,Harris2020}, SCIPY \citep{Virtanen2020}.}

\appendix
\section{Line profiles\label{appendix}}
\begin{figure*}
\includegraphics[width=\textwidth]{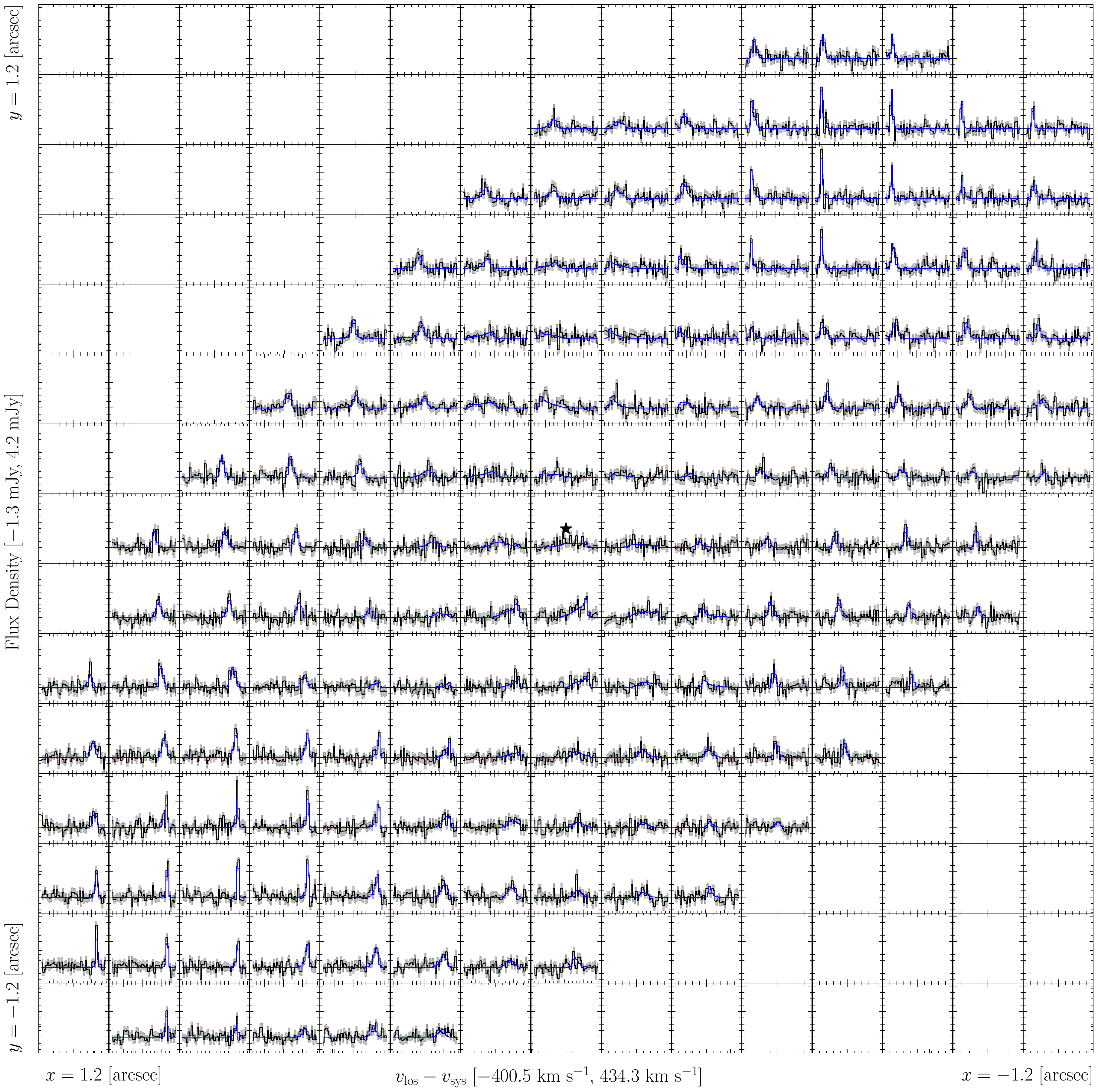}
\caption{Observed (black) and best-fit fiducial model (blue) line profiles in the fitting ellipse, on the down-sampled pixel scale, with noise per channel shown in gray.
The BH location is shown with a black star.
The x-axis is labeled with the velocity channel ranges for each panel, and the y-axis is labeled with the minimum and maximum flux shown in each panel.
The physical extent of the fitting region is also indicated at the corners of the x and y axes.
This figure indicates the model is a good fit to the observations across the gas disk.
}
\label{fig_fiducial_lps}
\end{figure*}

The model is fit to the data cube on the down-sampled pixel scale within the fitting ellipse, as discussed in \S\ref{model}, with $\chi^2$ calculated from the model and observed line profiles at each pixel.
The fiducial model and data cube line profiles for every down-sampled pixel in the ellipse are plotted in Figure \ref{fig_fiducial_lps}.

\begin{thebibliography}{40}
\bibitem[Astropy Collaboration et al.(2022)]{Astropy2022} Astropy Collaboration, Price-Whelan, A.~M., Lim, P.~L., et al.\ 2022, \apj, 935, 167. doi:10.3847/1538-4357/ac7c74

\bibitem[Astropy Collaboration et al.(2018)]{Astropy2018} Astropy Collaboration, Price-Whelan, A.~M., Sip{\H{o}}cz, B.~M., et al.\ 2018, \aj, 156, 123. doi:10.3847/1538-3881/aabc4f

\bibitem[Astropy Collaboration et al.(2013)]{Astropy2013} Astropy Collaboration, Robitaille, T.~P., Tollerud, E.~J., et al.\ 2013, \aap, 558, A33. doi:10.1051/0004-6361/201322068

\bibitem[Barth et al.(2001)]{Barth2001} Barth, A.~J., Sarzi, M., Rix, H.-W., et al.\ 2001, \apj, 555, 685

\bibitem[Barth et al.(2016)]{Barth2016a} Barth, A.~J., Boizelle, B.~D., Darling, J., et al.\ 2016, \apjl, 822, L28. doi:10.3847/2041-8205/822/2/L28

\bibitem[Barth et al.(2016)]{Barth2016b} Barth, A.~J., Darling, J., Baker, A.~J., et al.\ 2016, \apj, 823, 51. doi:10.3847/0004-637X/823/1/51

\bibitem[Beasley et al.(2018)]{Beasley2018} Beasley, M.~A., Trujillo, I., Leaman, R., et al.\ 2018, \nat, 555, 483

\bibitem[Binney \& Tremaine(2008)]{Binney2008} Binney, J. \& Tremaine, S.\ 2008, Galactic Dynamics: Second Edition, by James Binney and Scott Tremaine. ISBN 978-0-691-13026-2 (HB). Published by Princeton University Press, Princeton, NJ USA, 2008.

\bibitem[Bogd{\'a}n et al.(2024)]{Bogdan2024} Bogd{\'a}n, {\'A}., Goulding, A.~D., Natarajan, P., et al.\ 2024, Nature Astronomy, 8, 126. doi:10.1038/s41550-023-02111-9

\bibitem[Boizelle et al.(2017)]{Boizelle2017} Boizelle, B.~D., Barth, A.~J., Darling, J., et al.\ 2017, \apj, 845, 170

\bibitem[Boizelle et al.(2019)]{Boizelle2019} Boizelle, B.~D., Barth, A.~J., Walsh, J.~L., et al.\ 2019, \apj, 881, 10

\bibitem[Boizelle et al.(2021)]{Boizelle2021} Boizelle, B.~D., Walsh, J.~L., Barth, A.~J., et al.\ 2021, \apj, 908, 19. doi:10.3847/1538-4357/abd24d

\bibitem[Briggs(1995)]{Briggs1995} Briggs, D.~S.\ 1995, \aas

\bibitem[Buote \& Barth(2018)]{Buote2018} Buote, D.~A. \& Barth, A.~J.\ 2018, \apj, 854, 143. doi:10.3847/1538-4357/aaa971

\bibitem[Buote \& Barth(2019)]{Buote2019} Buote, D.~A. \& Barth, A.~J.\ 2019, \apj, 877, 91. doi:10.3847/1538-4357/ab1008

\bibitem[Cappellari(2002)]{Cappellari2002} Cappellari, M.\ 2002, \mnras, 333, 400

\bibitem[Carilli \& Walter(2013)]{Carilli2013} Carilli, C.~L. \& Walter, F.\ 2013, \araa, 51, 105

\bibitem[CASA Team et al.(2022)]{CASA2022} CASA Team, Bean, B., Bhatnagar, S., et al.\ 2022, \pasp, 134, 114501. doi:10.1088/1538-3873/ac9642

\bibitem[Cohn et al.(2021)]{Cohn2021} Cohn, J.~H., Walsh, J.~L., Boizelle, B.~D., et al.\ 2021, \apj, 919, 77. doi:10.3847/1538-4357/ac0f78

\bibitem[Cohn et al.(2023)]{Cohn2023} Cohn, J.~H., Curliss, M., Walsh, J.~L., et al.\ 2023, arXiv:2310.11296. doi:10.48550/arXiv.2310.11296

\bibitem[Davis et al.(2017)]{Davis2017} Davis, T.~A., Bureau, M., Onishi, K., et al.\ 2017, \mnras, 468, 4675. doi:10.1093/mnras/stw3217

\bibitem[Davis et al.(2018)]{Davis2018} Davis, T.~A., Bureau, M., Onishi, K., et al.\ 2018, \mnras, 473, 3818. doi:10.1093/mnras/stx2600

\bibitem[de Vaucouleurs et al.(1976)]{Vaucouleurs1976} de Vaucouleurs, G., de Vaucouleurs, A., \& Corwin, J.~R.\ 1976, Second reference catalogue of bright galaxies, 0

\bibitem[Dominiak et al.(2024a)]{Dominiak2024a} Dominiak, P., Bureau, M., Davis, T.~A., et al.\ 2024, \mnras, 529, 1597. doi:10.1093/mnras/stae314

\bibitem[Dominiak et al.(2024b)]{Dominiak2024b} Dominiak, P., Cappellari, M., Bureau, M., et al.\ 2024, arXiv:2404.11260. doi:10.48550/arXiv.2404.11260

\bibitem[Emsellem et al.(1994)]{Emsellem1994} Emsellem, E., Monnet, G., \& Bacon, R.\ 1994, \aap, 285, 723

\bibitem[Emsellem(2013)]{Emsellem2013} Emsellem, E.\ 2013, \mnras, 433, 1862. doi:10.1093/mnras/stt840

\bibitem[Faber et al.(1997)]{Faber1997} Faber, S.~M., Tremaine, S., Ajhar, E.~A., et al.\ 1997, \aj, 114, 1771. doi:10.1086/118606

\bibitem[Ferrarese \& Merritt(2000)]{Ferrarese2000} Ferrarese, L. \& Merritt, D.\ 2000, \apjl, 539, L9. doi:10.1086/312838

\bibitem[Ferr{\'e}-Mateu et al.(2015)]{Mateu2015} Ferr{\'e}-Mateu, A., Mezcua, M., Trujillo, I., et al.\ 2015, \apj, 808, 79

\bibitem[Ferr{\'e}-Mateu et al.(2017)]{Mateu2017} Ferr{\'e}-Mateu, A., Trujillo, I., Mart{\'\i}n-Navarro, I., et al.\ 2017, \mnras, 467, 1929

\bibitem[Fomalont et al.(2014)]{Fomalont2014} Fomalont, E., van Kempen, T., Kneissl, R., et al.\ 2014, The Messenger, 155, 19

\bibitem[Fonseca Alvarez et al.(2020)]{FonsecaAlvarez2020} Fonseca Alvarez, G., Trump, J.~R., Homayouni, Y., et al.\ 2020, \apj, 899, 73. doi:10.3847/1538-4357/aba001

\bibitem[Gebhardt et al.(2000)]{Gebhardt2000} Gebhardt, K., Bender, R., Bower, G., et al.\ 2000, \apjl, 539, L13. doi:10.1086/312840

\bibitem[Gonzaga et al.(2012)]{Gonzaga2012} Gonzaga, S., Hack, W., Fruchter, A., et al.\ 2012, The DrizzlePac Handbook, HST Data Handbook

\bibitem[Graham et al.(2016a)]{Graham2016a} Graham, A.~W., Durr{\'e}, M., Savorgnan, G.~A.~D., et al.\ 2016, \apj, 819, 43. doi:10.3847/0004-637X/819/1/43

\bibitem[Graham et al.(2016b)]{Graham2016b} Graham, A.~W., Ciambur, B.~C., \& Savorgnan, G.~A.~D.\ 2016, \apj, 831, 132. doi:10.3847/0004-637X/831/2/132

\bibitem[G{\"u}ltekin et al.(2009)]{Gultekin2009} G{\"u}ltekin, K., Richstone, D.~O., Gebhardt, K., et al.\ 2009, \apj, 698, 198. doi:10.1088/0004-637X/698/1/198

\bibitem[Hack et al.(2012)]{Hack2012} Hack, W.~J., Dencheva, N., Fruchter, A.~S., et al.\ 2012, \aas

\bibitem[Harris et al.(2020)]{Harris2020} Harris, C.~R., Millman, K.~J., van der Walt, S.~J., et al.\ 2020, \nat, 585, 357. doi:10.1038/s41586-020-2649-2

\bibitem[Hilz et al.(2013)]{Hilz2013} Hilz, M., Naab, T., \& Ostriker, J.~P.\ 2013, \mnras, 429, 2924. doi:10.1093/mnras/sts501

\bibitem[Hunter(2007)]{Hunter2007} Hunter, J.~D.\ 2007, Computing in Science and Engineering, 9, 90

\bibitem[Izumi et al.(2019)]{Izumi2019} Izumi, T., Onoue, M., Matsuoka, Y., et al.\ 2019, \pasj, 71, 111. doi:10.1093/pasj/psz096

\bibitem[Jeter et al.(2019)]{Jeter2019} Jeter, B., Broderick, A.~E., \& McNamara, B.~R.\ 2019, \apj, 882, 82. doi:10.3847/1538-4357/ab3221

\bibitem[Kabasares et al.(2022)]{Kabasares2022} Kabasares, K.~M., Barth, A.~J., Buote, D.~A., et al.\ 2022, \apj, 934, 162. doi:10.3847/1538-4357/ac7a38

\bibitem[Kabasares \& Cohn et al.(2024)]{KabasaresCohn2024} Kabasares, K.~M., Cohn, J.~H., Barth, A.~J., et al.\ 2024, \apj, 966, 132. doi:10.3847/1538-4357/ad2f36

\bibitem[Kang \& Lee(2021)]{Kang2021} Kang, J. \& Lee, M.~G.\ 2021, \apj, 914, 20. doi:10.3847/1538-4357/abf433

\bibitem[Kormendy \& Ho(2013)]{Kormendy2013} Kormendy, J., \& Ho, L.~C.\ 2013, \araa, 51, 511

\bibitem[Kormendy \& Richstone(1995)]{Kormendy1995} Kormendy, J. \& Richstone, D.\ 1995, \araa, 33, 581. doi:10.1146/annurev.aa.33.090195.003053

\bibitem[Krajnovi{\'c} et al.(2006)]{Krajnovic2006} Krajnovi{\'c}, D., Cappellari, M., de Zeeuw, P.~T., et al.\ 2006, \mnras, 366, 787

\bibitem[Krajnovi{\'c} et al.(2009)]{Krajnovic2009} Krajnovi{\'c}, D., McDermid, R.~M., Cappellari, M., et al.\ 2009, \mnras, 399, 1839. doi:10.1111/j.1365-2966.2009.15415.x

\bibitem[Krajnovi{\'c} et al.(2018)]{Krajnovic2018} Krajnovi{\'c}, D., Cappellari, M., McDermid, R.~M., et al.\ 2018, \mnras, 477, 3030. doi:10.1093/mnras/sty778

\bibitem[Krist \& Hook(2004)]{Krist2004} Krist, J., \& Hook, R. 2004, The Tiny Tim User’s Guide,
\url{http://www.stsci.edu/hst/observatory/focus/TinyTim}, Baltimore:
STScI

\bibitem[Kroupa(2001)]{Kroupa2001} Kroupa, P.\ 2001, \mnras, 322, 231. doi:10.1046/j.1365-8711.2001.04022.x

\bibitem[La Barbera et al.(2019)]{LaBarbera2019} La Barbera, F., Vazdekis, A., Ferreras, I., et al.\ 2019, \mnras, 489, 4090. doi:10.1093/mnras/stz2192

\bibitem[Larson et al.(2023)]{Larson2023} Larson, R.~L., Finkelstein, S.~L., Kocevski, D.~D., et al.\ 2023, \apjl, 953, L29. doi:10.3847/2041-8213/ace619

\bibitem[L{\"a}sker et al.(2014)]{Lasker2014} L{\"a}sker, R., Ferrarese, L., van de Ven, G., et al.\ 2014, \apj, 780, 70 

\bibitem[Lauer et al.(2007)]{Lauer2007} Lauer, T.~R., Faber, S.~M., Richstone, D., et al.\ 2007, \apj, 662, 808. doi:10.1086/518223

\bibitem[Lavezzi et al.(1999)]{Lavezzi1999} Lavezzi, T.~E., Dickey, J.~M., Casoli, F., et al.\ 1999, \aj, 117, 1995

\bibitem[Liddle(2007)]{Liddle2007} Liddle, A.~R.\ 2007, \mnras, 377, L74. doi:10.1111/j.1745-3933.2007.00306.x

\bibitem[Lucy(1974)]{Lucy1974} Lucy, L.~B.\ 1974, \aj, 79, 745

\bibitem[Maiolino et al.(2023)]{Maiolino2023} Maiolino, R., Scholtz, J., Witstok, J., et al.\ 2023, arXiv:2305.12492. doi:10.48550/arXiv.2305.12492

\bibitem[Marconi \& Hunt(2003)]{Marconi2003} Marconi, A. \& Hunt, L.~K.\ 2003, \apjl, 589, L21. doi:10.1086/375804

\bibitem[Mart{\'\i}n-Navarro et al.(2015a)]{Martin2015a} Mart{\'\i}n-Navarro, I., La Barbera, F., Vazdekis, A., et al.\ 2015, \mnras, 447, 1033. doi:10.1093/mnras/stu2480

\bibitem[Mart{\'\i}n-Navarro et al.(2015b)]{Martin2015} Mart{\'\i}n-Navarro, I., La Barbera, F., Vazdekis, A., et al.\ 2015, \mnras, 451, 1081, doi:10.1093/mnras/stv1022

\bibitem[Matt et al.(2023)]{Matt2023} Matt, C., G{\"u}ltekin, K., \& Simon, J.\ 2023, \mnras, 524, 4403. doi:10.1093/mnras/stad2146

\bibitem[McConnell \& Ma(2013)]{McConnellMa2013} McConnell, N.~J. \& Ma, C.-P.\ 2013, \apj, 764, 184

\bibitem[McMullin et al.(2007)]{McMullin2007} McMullin, J.~P., Waters, B., Schiebel, D., et al.\ 2007, Astronomical Data Analysis Software and Systems XVI, 376, 127

\bibitem[Mehrgan et al.(2024)]{Mehrgan2024} Mehrgan, K., Thomas, J., Saglia, R., et al.\ 2024, \apj, 961, 127. doi:10.3847/1538-4357/acfe09

\bibitem[Mould et al.(2000)]{Mould2000} Mould, J.~R., Huchra, J.~P., Freedman, W.~L., et al.\ 2000, \apj, 529, 786

\bibitem[Naab et al.(2009)]{Naab2009} Naab, T., Johansson, P.~H., \& Ostriker, J.~P.\ 2009, \apjl, 699, L178. doi:10.1088/0004-637X/699/2/L178

\bibitem[Nagai et al.(2019)]{Nagai2019} Nagai, H., Onishi, K., Kawakatu, N., et al.\ 2019, \apj, 883, 193. doi:10.3847/1538-4357/ab3e6e

\bibitem[Neumayer et al.(2007)]{Neumayer2007} Neumayer, N., Cappellari, M., Reunanen, J., et al.\ 2007, \apj, 671, 1329. doi:10.1086/523039

\bibitem[Nguyen et al.(2020)]{Nguyen2020} Nguyen, D.~D., den Brok, M., Seth, A.~C., et al.\ 2020, \apj, 892, 68

\bibitem[Nguyen et al.(2022)]{Nguyen2022} Nguyen, D.~D., Bureau, M., Thater, S., et al.\ 2022, \mnras, 509, 2920. doi:10.1093/mnras/stab3016

\bibitem[North et al.(2019)]{North2019} North, E.~V., Davis, T.~A., Bureau, M., et al.\ 2019, \mnras, 490, 319. doi:10.1093/mnras/stz2598

\bibitem[Onishi et al.(2017)]{Onishi2017} Onishi, K., Iguchi, S., Davis, T.~A., et al.\ 2017, \mnras, 468, 4663. doi:10.1093/mnras/stx631

\bibitem[Oser et al.(2010)]{Oser2010} Oser, L., Ostriker, J.~P., Naab, T., et al.\ 2010, \apj, 725, 2312. doi:10.1088/0004-637X/725/2/2312

\bibitem[Pacucci et al.(2023)]{Pacucci2023} Pacucci, F., Nguyen, B., Carniani, S., et al.\ 2023, \apjl, 957, L3. doi:10.3847/2041-8213/ad0158

\bibitem[Peng et al.(2010)]{Peng2010} Peng, C.~Y., Ho, L.~C., Impey, C.~D., et al.\ 2010, \aj, 139, 2097

\bibitem[Pensabene et al.(2020)]{Pensabene2020} Pensabene, A., Carniani, S., Perna, M., et al.\ 2020, \aap, 637, A84. doi:10.1051/0004-6361/201936634

\bibitem[Richardson(1972)]{Richardson1972} Richardson, W.~H.\ 1972, Journal of the Optical Society of America (1917-1983), 62, 55

\bibitem[Rieke \& Lebofsky(1985)]{Rieke1985} Rieke, G.~H. \& Lebofsky, M.~J.\ 1985, \apj, 288, 618. doi:10.1086/162827

\bibitem[Ruffa et al.(2019)]{Ruffa2019} Ruffa, I., Prandoni, I., Laing, R.~A., et al.\ 2019, \mnras, 484, 4239. doi:10.1093/mnras/stz255

\bibitem[Ruffa et al.(2023)]{Ruffa2023} Ruffa, I., Davis, T.~A., Cappellari, M., et al.\ 2023, \mnras, 522, 6170. doi:10.1093/mnras/stad1119

\bibitem[Rusli et al.(2011)]{Rusli2011} Rusli, S.~P., Thomas, J., Erwin, P., et al.\ 2011, \mnras, 410, 1223. doi:10.1111/j.1365-2966.2010.17610.x

\bibitem[Rusli et al.(2013)]{Rusli2013a} Rusli, S.~P., Thomas, J., Saglia, R.~P., et al.\ 2013, \aj, 146, 45

\bibitem[Rusli et al.(2013)]{Rusli2013b} Rusli, S.~P., Erwin, P., Saglia, R.~P., et al.\ 2013, \aj, 146, 160

\bibitem[Saglia et al.(2016)]{Saglia2016} Saglia, R.~P., Opitsch, M., Erwin, P., et al.\ 2016, \apj, 818, 47

\bibitem[Sandstrom et al.(2013)]{Sandstrom2013} Sandstrom, K.~M., Leroy, A.~K., Walter, F., et al.\ 2013, \apj, 777, 5

\bibitem[Savorgnan \& Graham(2016)]{SavorgnanGraham2016} Savorgnan, G.~A.~D. \& Graham, A.~W.\ 2016, \mnras, 457, 320. doi:10.1093/mnras/stv2713

\bibitem[Savorgnan et al.(2016)]{Savorgnan2016} Savorgnan, G.~A.~D., Graham, A.~W., Marconi, A., et al.\ 2016, \apj, 817, 21. doi:10.3847/0004-637X/817/1/21 

\bibitem[Scharw{\"a}chter et al.(2013)]{Scharwachter2013} Scharw{\"a}chter, J., McGregor, P.~J., Dopita, M.~A., et al.\ 2013, \mnras, 429, 2315. doi:10.1093/mnras/sts502

\bibitem[Scharw{\"a}chter et al.(2016)]{Scharwachter2016} Scharw{\"a}chter, J., Combes, F., Salom{\'e}, P., et al.\ 2016, \mnras, 457, 4272

\bibitem[Schlafly \& Finkbeiner(2011)]{Schlafly2011} Schlafly, E.~F. \& Finkbeiner, D.~P.\ 2011, \apj, 737, 103

\bibitem[Schulze \& Gebhardt(2011)]{Schulze2011} Schulze, A. \& Gebhardt, K.\ 2011, \apj, 729, 21. doi:10.1088/0004-637X/729/1/21

\bibitem[Seth et al.(2010)]{Seth2010} Seth, A.~C., Cappellari, M., Neumayer, N., et al.\ 2010, \apj, 714, 713. doi:10.1088/0004-637X/714/1/713

\bibitem[Shen et al.(2023)]{Shen2023} Shen, Y., Grier, C.~J., Horne, K., et al.\ 2023, arXiv:2305.01014. doi:10.48550/arXiv.2305.01014

\bibitem[Smith et al.(2019)]{Smith2019} Smith, M.~D., Bureau, M., Davis, T.~A., et al.\ 2019, \mnras, 485, 4359. doi:10.1093/mnras/stz625

\bibitem[Smith et al.(2021)]{Smith2021} Smith, M.~D., Bureau, M., Davis, T.~A., et al.\ 2021, \mnras, 503, 5984. doi:10.1093/mnras/stab791

\bibitem[Speagle(2020)]{Speagle2020} Speagle, J.~S.\ 2020, \mnras, 493, 3132. doi:10.1093/mnras/staa278

\bibitem[Trujillo et al.(2014)]{Trujillo2014} Trujillo, I., Ferr{\'e}-Mateu, A., Balcells, M., et al.\ 2014, \apjl, 780, L20

\bibitem[van den Bosch et al.(2012)]{Bosch2012} van den Bosch, R.~C.~E., Gebhardt, K., G{\"u}ltekin, K., et al.\ 2012, \nat, 491, 729

\bibitem[van den Bosch et al.(2015)]{Bosch2015} van den Bosch, R.~C.~E., Gebhardt, K., G{\"u}ltekin, K., et al.\ 2015, \apjs, 218, 10

\bibitem[van den Bosch(2016)]{Bosch2016} van den Bosch, R.~C.~E.\ 2016, \apj, 831, 134 

\bibitem[van der Marel \& van den Bosch(1998)]{MarelBosch1998} van der Marel, R.~P. \& van den Bosch, F.~C.\ 1998, \aj, 116, 2220. doi:10.1086/300593

\bibitem[van der Walt et al.(2011)]{Walt2011} van der Walt, S., Colbert, S.~C., \& Varoquaux, G.\ 2011, Computing in Science and Engineering, 13, 22

\bibitem[van der Walt et al.(2014)]{Walt2014}
van der Walt, S., Sch\"onberger, J., Nunez-Iglesias, J, et al.\ 2014, PeerJ, 2, e453

\bibitem[van der Wel et al.(2014)]{Wel2014} van der Wel, A., Franx, M., van Dokkum, P.~G., et al.\ 2014, \apj, 788, 28. doi:10.1088/0004-637X/788/1/28

\bibitem[van Dokkum et al.(2010)]{Dokkum2010} van Dokkum, P.~G., Whitaker, K.~E., Brammer, G., et al.\ 2010, \apj, 709, 1018

\bibitem[Vazdekis et al.(2010)]{Vazdekis2010} Vazdekis, A., S{\'a}nchez-Bl{\'a}zquez, P., Falc{\'o}n-Barroso, J., et al.\ 2010, \mnras, 404, 1639

\bibitem[Viaene et al.(2017)]{Viaene2017} Viaene, S., Sarzi, M., Baes, M., et al.\ 2017, \mnras, 472, 1286. doi:10.1093/mnras/stx1781

\bibitem[Virtanen et al.(2020)]{Virtanen2020} Virtanen, P., Gommers, R., Oliphant, T.~E., et al.\ 2020, Nature Methods, 17, 261. doi:10.1038/s41592-019-0686-2

\bibitem[Walsh et al.(2010)]{Walsh2010} Walsh, J.~L., Barth, A.~J., \& Sarzi, M.\ 2010, \apj, 721, 762

\bibitem[Walsh et al.(2013)]{Walsh2013} {Walsh}, J.~L., {Barth}, A.~J., {Ho}, L.~C, \& {Sarzi}, M.\ 2013, \apj, 770, 86

\bibitem[Walsh et al.(2015)]{Walsh2015} Walsh, J.~L., van den Bosch, R.~C.~E., Gebhardt, K., et al.\ 2015, \apj, 808, 183. doi:10.1088/0004-637X/808/2/18

\bibitem[Walsh et al.(2016)]{Walsh2016} Walsh, J.~L., van den Bosch, R.~C.~E., Gebhardt, K., et al.\ 2016, \apj, 817, 2

\bibitem[Walsh et al.(2017)]{Walsh2017} Walsh, J.~L., van den Bosch, R.~C.~E., Gebhardt, K., et al.\ 2017, \apj, 835, 208

\bibitem[Waters et al.(2024)]{Waters2024} Waters, T.~K., G{\"u}ltekin, K., Gebhardt, K., et al.\ 2024, arXiv:2406.14623

\bibitem[Wellons et al.(2016)]{Wellons2016} Wellons, S., Torrey, P., Ma, C.-P., et al.\ 2016, \mnras, 456, 1030

\bibitem[Willmer(2018)]{Willmer2018} Willmer, C.~N.~A.\ 2018, \apjs, 236, 47

\bibitem[Wilman et al.(2005)]{Wilman2005} Wilman, R.~J., Edge, A.~C., \& Johnstone, R.~M.\ 2005, \mnras, 359, 755. doi:10.1111/j.1365-2966.2005.08956.x

\bibitem[Y{\i}ld{\i}r{\i}m et al.(2015)]{Yildirim2015} Y{\i}ld{\i}r{\i}m, A., van den Bosch, R.~C.~E., van de Ven, G., et al.\ 2015, \mnras, 452, 1792. doi:10.1093/mnras/stv1381

\bibitem[{Y\i ld\i r\i m {et~al.}(2017)Y\i ld\i r\i m, Bosch, Ven, Mart\'{i}n-Navarro, Walsh, Husemann, G\"{u}ltekin, \& Gebhardt}]{Yildirim2017}
Y\i ld\i r\i m, A., van den Bosch, R.~E.~C., van den Ven, G., {et~al.} 2017, \mnras, 468, L4216

\bibitem[Zhu et al.(2021)]{Zhu2021} Zhu, P., Ho, L.~C., \& Gao, H.\ 2021, \apj, 907, 6. doi:10.3847/1538-4357/abcaa1

\end{thebibliography}
\end{document}